\documentclass[prd,showpacs,twocolumn,nofootinbib,eqsecnum]{revtex4}

\usepackage{amsmath,amsfonts,amssymb}
\usepackage{graphics}

\allowdisplaybreaks

\newcommand{\be}{\begin{equation}}
\newcommand{\ee}{\end{equation}}
\newcommand{\bea}{\begin{eqnarray}}
\newcommand{\eea}{\end{eqnarray}}
\newcommand{\ben}{\begin{eqnarray*}}
\newcommand{\een}{\end{eqnarray*}}

\newcommand{\pa}{\partial}

\newcommand{\vek}[1]{{\mathbf #1}}

\newcommand{\onunder}[2]{\genfrac{}{}{0pt}{}{#1}{#2}}

\begin{document}

\author{Christian K\"onigsd\"orffer}
\email{C.Koenigsdoerffer@tpi.uni-jena.de}
\affiliation{Theoretisch-Physikalisches Institut,
Friedrich-Schiller-Universit\"at\\ Max-Wien-Platz 1, 07743 Jena,
Germany}

\author{Guillaume Faye}
\affiliation{Theoretisch-Physikalisches Institut,
Friedrich-Schiller-Universit\"at\\ Max-Wien-Platz 1, 07743 Jena,
Germany}

\author{Gerhard Sch\"afer}
\affiliation{Theoretisch-Physikalisches Institut,
Friedrich-Schiller-Universit\"at\\ Max-Wien-Platz 1, 07743 Jena,
Germany}

\date{\today}

\title{The binary black-hole dynamics at the third-and-a-half
post-Newtonian order in the ADM-formalism
}
\begin{abstract}
We specialize the radiation-reaction part of the
Arnowitt-Deser-Misner (ADM)
Hamiltonian for many non-spinning point-like bodies, calculated by
Jaranowski and Sch\"afer \cite{1997Jaranowski}, to third-and-a-half
post-Newtonian approximation to general relativity, to binary systems.
This Hamiltonian is used for the computation of the instantaneous
gravitational energy loss of a binary to $1$PN reactive order.
We also derive the equations of motion, which include PN reactive terms
via Hamiltonian and Euler-Lagrangian approaches. The results are
consistent with the expressions for reactive acceleration provided by
Iyer-Will formalism in Ref.~\cite{IW95} in a general class of
gauges.
\end{abstract}


\maketitle

\section{Introduction}

Compact binaries are the most promising sources of detectable
gravitational waves for ground and space based interferometric
experiments, TAMA$300$, GEO$600$, VIRGO, LIGO (Laser Interferometer
Gravitational Wave Observatory) and LISA (Laser Interferometer Space
Antenna). Because of gravitational wave emission, the binary separation
decreases adiabatically in time while the frequency
increases according to the relativistic version of the Kepler law. This process
ends as soon as the first instable orbit is reached before the
coalescence. During the inspiraling phase the binary
system produces a most likely detectable signal.
The construction of wave patterns for data analysis
requires the knowledge of the theoretical wave form to very high
post-Newtonian orders [see, e.g., Refs.~\cite{DIS2001,DIS2002}]. This assumes
a good understanding of the compact-binary dynamics. The secular effects
governing the long-time evolution are of particular importance. They
result from the reaction force due to the quadrupole damping arising at
the $2.5$ order in power of $v^2/c^2$, $v$ representing a typical
orbital velocity and $c$ standing for the speed of light in vacuum. For
this reason, they are often referred to as the $2.5$ post-Newtonian (PN)
damping. The measurements resulting from the observation of the
Hulse-Taylor binary pulsar PSR $1913$+$16$ show excellent agreement
with the values predicted by Einstein theory~\cite{WT2002}.

A while ago, Iyer and Will \cite{IW93} derived the equations of motion
for two non-spinning point-like objects including the first dissipative
terms appearing at the 2.5 and 3.5PN level, \emph{i.e.} the quadrupole
damping and its 1PN correction. They proceeded by postulating a balance
equation between the instantaneous flux of energy and angular momentum
in the far zone on the one hand, and the instantaneous loss of energy
and angular momentum in the system's near zone on the other hand. In a
later paper \cite{IW95}, they detailed their derivation and proposed an
alternative method, based on the specialization of the Blanchet's
radiation-reaction potentials to binary systems, which led to
the same results \cite{B93}. More recently, the dynamics of a two-body
system at the $3.5$PN order was derived by Pati
and Will in harmonic coordinates \cite{PatiWill2002} using
the integration method for the Einstein field
equations developed in Ref.~\cite{PatiWill2000}.

This article deals with the $3.5$PN gravitational damping of compact
binaries regarded as point-masses. We base our investigation on the
$N$-body point-mass Hamiltonian \cite{1997Jaranowski} calculated by
Jaranowski and Sch\"afer within the Arnowitt-Deser-Misner (ADM)
Hamiltonian formalism of general relativity \cite{ADM62}, which has
proved to be very efficient in determining the approximate $3.5$PN
dynamics
\cite{1997Jaranowski,OOKH74-1598,S85,DamourJaranowskiSchaefer2000}.
A remarkable feature of the ADM formalism at this level is
the absence of asymptotic matching between
the near field and the far field given in a near-zone and a
far-zone-defined coordinate system respectively. We can stay in one
single global coordinate system in all our calculations, like in
Ref.~\cite{PatiWill2002}. Let us mention that the ADM formalism does
not provide directly the balance equations between instantaneous losses
and fluxes, even when assuming quasi-stationarity in the radiation
emission. Although the ADM formalism gives us the conserved total
energy, the balance between the lost energy of the matter system
and the energy flux in the wave zone is proved up to $3.5$PN
order only [see, e.g., Ref.~\cite{GS97}].

The main purposes of this work are \emph{(i)} to specialize the $N$-body
$3.5$PN ADM Hamiltonian of Ref.~\cite{1997Jaranowski} to $N=2$,
\emph{(ii)} to deduce from it the energy loss of a two-body system in
general orbits, and
\emph{(iii)} to calculate the equations of motions in the Hamiltonian
form and in the Euler-Lagrangian form in order to determine the
radiation-reaction force. The expression obtained for the
energy loss is different from that derived in Jaranowski and Sch\"afer
\cite{1997Jaranowski} and in Iyer and Will \cite{IW95}, but
the time-averaged expressions coincide in the case of quasi-elliptic
orbital motion. The plan of the paper is to reduce the
original Hamiltonian for two point-masses. Next, we perform a partial
time differentiation to get the energy loss. On the other side, we compute
the radiation-reaction force from the Euler-Lagrangian equations of
motion. The corresponding energy loss turns out to be equivalent to the
preceding one after averaging. Our results are compared with those of
Iyer and Will and the accessory gauge coefficients are determined
explicitly. Hadamard regularization is systematically employed.

We use units in which $16\pi G=1=c$, where $G$ is the Newtonian
gravitational constant, but $G$ and $c$ are restored in the presentation
of our final results. In our notation, $t$ is the coordinate time.
Characters in bold, like $\vek{x}=\left(x^i\right)$, represents a
point in the $3$-dimensional Euclidean space $\mathbb{R}^3$ endowed with a
standard Euclidean metric. The scalar product is denoted by a
dot. Letters $a,b,\ldots$ are particle labels, so that $\vek{x}_a \in
\mathbb{R}^3$ denotes the position of the $a$th particle. We also
define: $\vek{r}_a := \vek{x} - \vek{x}_a$, $r_a := |\vek{r}_a|$,
$\vek{n}_a := \vek{r}_a/r_a$; and for
$a\ne b$: $\vek{r}_{ab}:=\vek{x}_a -\vek{x}_b =\vek{r}_b - \vek{r}_a$,
$r_{ab} :=|\vek{r}_{ab}|$,
$\vek{n}_{ab} := \vek{r}_{ab}/r_{ab}$; $|\ldots|$
stands here for the Euclidean length of a vector. The momentum vector of
the particle $a$ with mass parameter $m_a$ is denoted by
$\vek{p}_a=\left(p_{ai}\right)$. A dot over a symbol, like in
$\dot{\vek{x}}_a$, represents its total time derivative. The partial
differentiation with respect to $x^i$ is denoted by $\pa_i$, or
equivalently by a comma, \emph{i.e.} $\pa_i\phi\equiv\phi_{,i}$.

\section{$\mathbf{3.5}$PN field equations}

The ADM formulation is based on a $(3+1)$ splitting of space-time. The
canonical variables are, on one hand, the projection $g_{ij}$ of the
$3$-metric on the space-like hypersurfaces $x^0 = t =\text{const.}$ in
a suitable coordinate grid and, on the other hand, its conjugate momentum
$\pi^{ij}$. In the so-called ADM gauge \cite{OOKH74-1598,S85,DS88}, the
spatial metric $g_{ij}$ decomposes into a diagonal part $\left( 1 +
\frac{1}{8} \phi \right)^4 \delta_{ij}$, plus a part $h_{ij}^\text{TT}$
transverse and traceless with respect to the Euclidean metric
$\delta_{ij}$, so that
\be
g_{ij}=\left(1+\frac{1}{8}\phi\right)^4 \delta_{ij} + h_{ij}^\text{TT}
\ee
with $\partial_j h_{ij}^\text{TT} = 0$. In addition, the field
momentum $\pi^{ij}$ is traceless: $\pi^{ii} = 0$. Let us consider now
$h_{ij}^\text{TT}$ as the new field variable, and denote its conjugate
momentum by $\pi^{ij\text{TT}}$. The degrees of freedom associated with
$\tilde{\pi}^{ij} = \pi^{ij} - \pi^{ij\text{TT}}$ and the potential
$\phi$ are removed by solving the constraint equations.

After inserting the values of $\tilde{\pi}^{ij}$ and $\phi$ into the
general relativistic Hamiltonian of the isolated system under
consideration in an
asymptotically flat space-time, the latter simplifies to a
surface integral often referred to as the reduced Hamiltonian $H$. It
is a functional of the independent degrees of freedom $h_{ij}^\text{TT}$
and $\pi^{ij\text{TT}}$ \cite{ADM62}.

For a system of $N$ point-like bodies with position vectors $\vek{x}_a$
and momenta $\vek{p}_a$ ($a=1,\ldots,N$), the Hamiltonian takes the form
\be
\label{eq:full_hamiltonian}
H = H
\left[\vek{x}_a,\vek{p}_a,h_{i j}^{\rm TT},\pi^{ij{\rm TT}}\right]
\,.
\ee
The Hamiltonian equations of motion for body $a$ are
\be
\label{eq:HamBewegungsGlDefinition}
\dot {\vek p}_a =-\frac{\partial H}{\partial \vek{x}_a}
\quad \quad {\rm and} \quad \quad
\dot {\vek x}_a =\frac{\partial H}{\partial \vek{p}_a}\,,
\ee
and the field equations for the independent degrees of freedom read
\be
\frac{\pa}{\pa t}\pi^{ij{\rm TT}} = -\delta^{{\rm TT}ij}_{kl}
\frac{\delta H}{\delta h^{\rm TT}_{kl}}
\,,\quad
\frac{\pa}{\pa t} h^{\rm TT}_{ij} = \delta^{{\rm TT}kl}_{ij}
\frac{\delta H}{\delta \pi^{kl{\rm TT}}}
\,,
\label{eq:eomgf}
\ee
where $(\delta\ldots)/(\delta\ldots)$ denotes the Fr\'echet
derivative [see, e.g., Ref.~\cite{Schwartz1992}] and
where $\delta^{{\rm TT}ij}_{kl}$ is the TT-projection
operator [see, e.g., Eq.~(2.17) of \cite{S85}].

The Hamiltonian $H$ can be uniquely decomposed into a matter part,
$H^{\rm mat}$, depending on the matter variables only, a field part,
$H^{\rm field}$, depending on the field variables only, and an
interaction part, $H^{\rm int}$, depending on both sets of variables,
matter and field, in a way that it vanishes if one set is put to
zero. In short, the full content of the field-plus-matter dynamics
at the $3.5$PN order is included in the Hamiltonian
\be
\label{eq:ha}
H_{\le{\rm 3.5PN}} =
H_{\le{\rm 3.5PN}}^{\rm mat}
+H_{\le{\rm 3.5PN}}^{\rm field}
+H_{\le{\rm 3.5PN}}^{\rm int}
\,.
\ee
The notation $\le3.5$PN indicates that all orders lower or equal than
$3.5$PN are included. 
In what follows, only the interaction part of the
Hamiltonian~\eqref{eq:ha} will be needed~\cite{S85}. It has been explicitly
computed at the dissipative $3.5$PN order in Ref.~\cite{1997Jaranowski} by
Jaranowski and Sch\"afer. Its truncation at the $2.5$PN level,
$H_{{\rm 2.5PN}}^{\rm int}$, takes the well-known form presented in
Refs.~\cite{S85,S95}
\be
\label{eq:25res}
H_{{\rm 2.5PN}}^{\rm int} \left(\vek{x}_a,\vek{p}_a,t\right) =
5\pi\, \dot{\chi}_{(4)ij}(t)\, \chi_{(4)ij}\left(\vek{x}_a,\vek{p}_a\right)
\ee
with 
\begin{align}
\chi_{(4)ij} \left(\vek{x}_a,\vek{p}_a\right) :=\, &\frac{1}{60\pi}
\bigg[
\sum\limits_a \frac{2}{m_a}
\left( \vek{p}_a^2 \delta_{ij} - 3 p_{ai} p_{aj} \right)
\nonumber \\
&+\frac{1}{16\pi}
\sum\limits_a \sum\limits_{b\ne a} \frac{m_a m_b}{r_{ab}}
\left( 3 n_{ab}^i n_{ab}^j - \delta_{ij} \right)
\bigg]
\,.
\label{eq:chidef}
\end{align}
The function $\dot{\chi}_{(4)ij}(t)$ comes from the post-Newtonian
expansion of the retarded integral giving $h_{i j}^{\rm TT}$ as a
function of $\vek{x}_{a}$ and $\vek{p}_{a}$~\cite{1997Jaranowski}.
As $\partial H /\partial \vek{x}_a$ and
$\partial H /\partial \vek{p}_a$ are differentiated for constant 
$h_{i j}^{\rm TT}$ and $\pi^{ij{\rm TT}}$, the positions and momenta
appearing in  Eq.~\eqref{eq:chidef} are not affected by the operators
$\partial /\partial \vek{x}_a$ and
$\partial /\partial \vek{p}_a$. In order not to confuse the
latter particle variables with the ones that are already present in the
original Hamiltonian~\eqref{eq:full_hamiltonian} and are affected by
$\partial /\partial \vek{x}_a$ and
$\partial /\partial \vek{p}_a$, we shall mark them with a prime
symbol: $\vek{x}_{a'}$, $\vek{p}_{a'}$. Thus, in our notation,
$\dot{\chi}_{(4)ij}(t)$ denotes the time derivative of
$\chi_{(4)ij}(\vek{x}_{a'}, \vek{p}_{a'})$. 

\begin{widetext}
The $3.5$PN Hamiltonian $H_{{\rm 3.5PN}}^{\rm int}$
reads\footnote{We observe that there are few misprints in the
expressions for $H_{{\rm 3.5PN}}^{\rm int}$, appearing
in~\cite{1997Jaranowski}. We make appropriate changes in those
expressions and their positions are marked
by~\fbox{\rule[0.1cm]{0cm}{0.1cm}$\ldots$}.}
\begin{align}
\label{eq:35res}
H_{{\rm 3.5PN}}^{\rm int}\left(\vek{x}_a,\vek{p}_a,t\right)=\,
&5\pi\,\chi_{(4)ij} \left(\vek{x}_a,\vek{p}_a\right)
\left[\dot{\Pi}_{1ij}\left(t\right)+\dot{\Pi}_{2ij}\left(t\right)
+\fbox{$\ddot{\Pi}_{3ij}\left(t\right)$}\; \right] + 5\pi\,
\fbox{$\dot{\chi}_{(4)ij}\left(t\right)$}
\left[{\Pi}_{1ij}\left(\vek{x}_a,\vek{p}_a\right)
+\widetilde{\Pi}_{2ij}\left(\vek{x}_a,t\right) \right]
\nonumber\\
&-5\pi\,\ddot{\chi}_{(4)ij}\left(t\right) {\Pi}_{3ij}
\left(\vek{x}_a,\vek{p}_a\right)
+\dot{\chi}_{(4)ij}\left(t\right)
\Big[ {Q'}_{ij} \left(\vek{x}_a,\vek{p}_a,t\right)
+{Q''}_{ij} \left(\vek{x}_a,t\right) \Big]
\nonumber\\ &
+\frac{\pa^3}{\pa t^3}
\Big[{R'} \left(\vek{x}_a,\vek{p}_a,t\right)
+{R''} \left(\vek{x}_a,t\right) \Big]
\,.
\end{align}
$H_{{\rm 3.5PN}}^{\rm int}$ is parametrized by the following functions
\begin{subequations}
\label{H35_is_parametrized_by_the_functions}
\begin{align}
\Pi_{1ij}\left(\vek{x}_a,\vek{p}_a\right):=\,
&\frac{4}{15}\left(\frac{1}{16\pi}\right)
\sum\limits_a\frac{\vek{p}_a^2}{m_a^3}
\left( - \vek{p}_a^2\delta_{ij} + 3p_{ai}p_{aj} \right)
+\frac{8}{5}\left(\frac{1}{16\pi}\right)^2
\sum\limits_a\sum\limits_{b\ne a}
\frac{m_b}{m_a r_{ab}} \left( - 2 \vek{p}_a^2\delta_{ij} + 5p_{ai}p_{aj}
+ \vek{p}_a^2 n_{ab}^i n_{ab}^j \right)
\nonumber\\[1ex]
&+\frac{1}{5} \left(\frac{1}{16\pi}\right)^2
\sum\limits_a\sum\limits_{b\ne a} \frac{1}{r_{ab}} \bigg\{
\Big[19 (\vek{p}_a\cdot\vek{p}_b)- 3 (\vek{n}_{ab}\cdot\vek{p}_a)
(\vek{n}_{ab}\cdot\vek{p}_b)\Big] \delta_{ij} - 42 {p_{ai}p_{bj}}
\nonumber\\[1ex]
&\left. -3\Big[5(\vek{p}_a\cdot\vek{p}_b)
+ (\vek{n}_{ab}\cdot\vek{p}_a) (\vek{n}_{ab}\cdot\vek{p}_b) \Big]
n_{ab}^i n_{ab}^j\right.
+6 (\vek{n}_{ab}\cdot\vek{p}_b)
\left( n_{ab}^i p_{aj} + n_{ab}^j p_{ai} \right) \bigg\}
\nonumber\\[1ex]
&+\frac{41}{15} \left(\frac{1}{16\pi}\right)^3
\sum\limits_a\sum\limits_{b\ne a} \frac{m_a^2 m_b}{r_{ab}^2}
\left( \delta_{ij} - 3 n_{ab}^i n_{ab}^j \right)
\nonumber\\[1ex]
&+ \frac{1}{45} \left(\frac{1}{16\pi}\right)^3
\sum\limits_a\sum\limits_{b\ne a}\sum\limits_{c\ne a,b} m_a m_b m_c
\left\{ \frac{18}{r_{ab}r_{ca}}
\left( \delta_{ij} - 3 n_{ab}^i n_{ab}^j \right) \right.
\nonumber\\[1ex]
&- \frac{180}{s_{abc}} \left[
\left( \frac{1}{r_{ab}} + \frac{1}{s_{abc}} \right) n_{ab}^i n_{ab}^j
+ \frac{1}{s_{abc}} n_{ab}^i n_{bc}^j \right]
\nonumber\\[1ex]
&\left.+ \frac{10}{s_{abc}} \left[ 4 \left(
\frac{1}{r_{ab}} + \frac{1}{r_{bc}} + \frac{1}{r_{ca}} \right)
- \frac{r_{ab}^2 + r_{bc}^2 + r_{ca}^2}{r_{ab}r_{bc}r_{ca}} \right]\delta_{ij}
\right\}
\label{eq:P1res}
\,,
\end{align}
with $s_{abc}:=r_{ab}+r_{bc}+r_{ca}$,
\begin{align}
\Pi_{2ij}\left(\vek{x}_a,\vek{p}_a\right):=\,
&\frac{1}{5}\left(\frac{1}{16\pi}\right)^2
\sum\limits_a\sum\limits_{b\ne a} \frac{m_b}{m_a r_{ab}} \left\{
\Big[5(\vek{n}_{ab}\cdot\vek{p}_a)^2 - \vek{p}_a^2 \Big] \delta_{ij}
- 2p_{ai}p_{aj}
+ \Big[5\vek{p}_a^2 - 3(\vek{n}_{ab}\cdot\vek{p}_a)^2 \Big] n_{ab}^i
n_{ab}^j
\right.
\nonumber\\[1ex]
& \left. - 6(\vek{n}_{ab}\cdot\vek{p}_a)(n_{ab}^i p_{aj} + n_{ab}^j p_{ai})
\right\} 
+\frac{6}{5} \left(\frac{1}{16\pi}\right)^3
\sum\limits_a\sum\limits_{b\ne a} \frac{m_a^2 m_b}{r_{ab}^2}
\left( 3 n_{ab}^i n_{ab}^j - \delta_{ij} \right)
\nonumber\\[1ex]
&+\frac{1}{10}\left(\frac{1}{16\pi}\right)^3
\sum\limits_a\sum\limits_{b\ne a}\sum\limits_{c\ne a,b} m_a m_b m_c
\left\{
\left[
5\frac{r_{ca}}{r_{ab}^3}\left(1-\frac{r_{ca}}{r_{bc}}\right)
+\frac{13}{r_{ab}r_{ca}} -\frac{40}{r_{ab}s_{abc}}\right]\delta_{ij}
\right.
\nonumber\\[1ex]
&+\left[
3\frac{r_{ab}}{r_{ca}^3}
+\frac{r_{bc}^2}{r_{ab}r_{ca}^3}
-\frac{5}{r_{ab}r_{ca}} 
+\frac{40}{s_{abc}}\left(\frac{1}{r_{ab}}+\frac{1}{s_{abc}}\right)
\right] n_{ab}^i n_{ab}^j
\nonumber\\[1ex]
&\left.+\left[2 \frac{(r_{ab} + r_{ca})}{r_{bc}^3}
-16 \left( \frac{1}{r_{ab}^2} + \frac{1}{r_{ca}^2} \right)
+ \frac{88}{s_{abc}^2}
\right] n_{ab}^i n_{ca}^j
\right\}
\label{eq:P2res}
\,,
\end{align}
\begin{align}
\Pi_{3ij}\left(\vek{x}_a,\vek{p}_a\right):= 
\frac{1}{5} \left(\frac{1}{16\pi}\right)^2
\sum\limits_a\sum\limits_{b\ne a} m_b \Big[
- 5 (\vek{n}_{ab}\cdot\vek{p}_a) \delta_{ij}
+ (\vek{n}_{ab}\cdot\vek{p}_a) n_{ab}^i n_{ab}^j
+ 7 \left( n_{ab}^i p_{aj} + n_{ab}^j p_{ai} \right) \Big]
\label{eq:P3res}
\,,
\end{align}
\begin{align}
\widetilde{\Pi}_{2ij}\left(\vek{x}_a,t\right):=
&\mbox{}\frac{1}{5}\left(\frac{1}{16\pi}\right)^2
\sum\limits_{a}\sum\limits_{a'}\frac{m_a}{m_{a'}r_{aa'}}
\bigg\{
\Big[5(\vek{n}_{a{a'}}\cdot\vek{p}_{a'})^2-\vek{p}_{a'}^2 \Big]\delta_{ij}
-2 p_{{a'}i} p_{{a'}j}
\nonumber\\[1ex]
&+\Big[5\vek{p}_{a'}^2 - 3 (\vek{n}_{{aa'}}\cdot\vek{p}_{a'})^2 \Big]
n_{{aa'}}^i n_{{aa'}}^j
- 6 (\vek{n}_{{aa'}}\cdot\vek{p}_{a'})
\left( n_{aa'}^i p_{{a'}j} + n_{aa'}^j p_{{a'}i} \right)
\bigg\}
\nonumber\\[1ex]
&+\frac{1}{10}\left(\frac{1}{16\pi}\right)^3
\sum\limits_{a}\sum\limits_{a'}\sum\limits_{{b'}\ne {a'}} m_a m_{a'} m_{b'}
\Bigg\{
\frac{32}{s_{a{a'}{b'}}}
\left(\frac{1}{r_{{a'}{b'}}} + \frac{1}{s_{a{a'}{b'}}}\right)
n_{{a'}{b'}}^i n_{{a'}{b'}}^j
\nonumber\\[1ex]
&+ 16 \left( \frac{1}{r_{{a'}{b'}}^2} - \frac{2}{s_{a{a'}{b'}}^2} \right)
\left( n_{aa'}^i n_{{a'}{b'}}^j + n_{aa'}^j n_{{a'}{b'}}^i \right)
- 2 \left(\frac{r_{aa'} + r_{ab'}}{r_{{a'}{b'}}^3}
+ \frac{12}{s_{a{a'}{b'}}^2}\right) n_{aa'}^i n_{a{b'}}^j
\nonumber\\[1ex]
&+\left[
\frac{r_{aa'}}{r_{{a'}{b'}}^3} \left(\frac{r_{aa'}}{r_{ab'}} + 3\right)
-\frac{5}{r_{{a'}{b'}}r_{aa'}}
+\frac{8}{s_{a{a'}{b'}}}
\left(\frac{1}{r_{aa'}} + \frac{1}{s_{a{a'}{b'}}}\right)
\right] n_{aa'}^i n_{{aa'}}^j
\nonumber\\[1ex]
&+\left[
5\frac{r_{aa'}}{r_{{a'}{b'}}^3}
\left( 1 - \frac{r_{aa'}}{r_{ab'}} \right)
+ \frac{17}{r_{{a'}{b'}}r_{aa'}} - \frac{4}{r_{aa'} r_{ab'}}
- \frac{8}{s_{a{a'}{b'}}}
\left( \frac{1}{r_{aa'}} + \frac{4}{r_{{a'}{b'}}} \right)
\right]\delta_{ij}
\Bigg\}\,,
\label{eq:Pi2Schlange}
\end{align}
with $s_{aa'b'} := r_{aa'} + r_{ab'} + r_{a'b'}$. Variables originating from
$h_{ij}^\text{TT}$ and $\pi^{ij\text{TT}}$ are primed. It is indeed crucial to
distinguish the particle positions and momenta inside and outside the
transverse-traceless (TT) quantities at this stage.

The other four functions which enter $H_{{\rm 3.5PN}}^{\rm int}$ are
\begin{align}
{Q'}_{ij}\left(\vek{x}_a,\vek{p}_a,t\right):=
-\frac{1}{16}\left(\frac{1}{16\pi}\right)
\sum\limits_{a}\sum\limits_{a'}\frac{m_{a'}}{m_{a} r_{aa'}}
\Big[2p_{ai}p_{aj}+12(\vek{n}_{aa'}\cdot\vek{p}_a)n_{aa'}^i p_{aj}
-5 \vek{p}_a^2 n_{aa'}^i n_{aa'}^j
+ 3(\vek{n}_{aa'}\cdot \vek{p}_a)^2 n_{aa'}^i n_{aa'}^j \Big]\,,
\label{eq:Qstrich}
\end{align}
\begin{align}
{Q''}_{ij} \left(\vek{x}_a,t\right) :=
&\mbox{}\frac{1}{32} \left(\frac{1}{16\pi}\right)^2
\sum\limits_{a}\sum\limits_{b\ne a}\sum\limits_{a'} m_a m_b m_{a'}
\left\{
\frac{32}{s_{aba'}} \left( \frac{1}{r_{ab}} + \frac{1}{s_{aba'}}
\right) n_{ab}^i n_{ab}^j 
+\left[ 3 \frac{r_{aa'}}{r_{ab}^3} - \frac{5}{r_{ab}r_{aa'}}
+\frac{r_{ba'}^2}{r_{ab}^3r_{aa'}}
\right.
\right.
\nonumber\\[1ex]&
\left.
+ \frac{8}{s_{aba'}} \left( \frac{1}{r_{aa'}}
+ \frac{1}{s_{aba'}} \right) \right]
n_{aa'}^in_{aa'}^j
-2 \left( \frac{r_{aa'}+r_{ba'}}{r_{ab}^3} + \frac{12}{s_{aba'}^2} \right)
n_{aa'}^i n_{ba'}^j
\left.\fbox{$-$}\, 32 \left( \frac{1}{r_{ab}^2} - \frac{2}{s_{aba'}^2} \right)
n_{ab}^i n_{aa'}^j
\right\}
\,,
\label{eq:Qbis}
\end{align}
with $s_{aba'}:=r_{ab}+r_{aa'}+r_{ba'}$,
\begin{align}
{R'} \left(\vek{x}_a,\vek{p}_a,t\right):=
&\mbox{}\frac{2}{105} \left(\frac{1}{16\pi}\right)
\sum\limits_{a}\sum\limits_{a'} \frac{r_{aa'}^2}{m_{a}m_{a'}}
\Big[
- 5 \vek{p}_{a}^2 \vek{p}_{a'}^2
+ 11 (\vek{p}_{a}\cdot\vek{p}_{a'})^2 + 4(\vek{n}_{aa'}\cdot\vek{p}_{a'})^2
\vek{p}_{a}^2 
\nonumber\\[1ex]
&+4(\vek{n}_{aa'}\cdot\vek{p}_{a})^2 \vek{p}_{a'}^2 - 
12 (\vek{n}_{aa'}\cdot\vek{p}_{a'})
(\vek{n}_{aa'}\cdot\vek{p}_{a}) (\vek{p}_{a}\cdot\vek{p}_{a'}) \Big]
\nonumber\\[1ex]
&- \frac{1}{105} \left(\frac{1}{16\pi}\right)^2
\sum\limits_{a}\sum\limits_{a'}\sum\limits_{b'\ne a'}
\frac{m_{a'}m_{b'}}{m_a} \left[
\left( 2 \frac{r_{aa'}^4}{r_{a'b'}^3} - 2 \frac{r_{aa'}^2
r_{ab'}^2}{r_{a'b'}^3} - 5 \frac{r_{aa'}^2}{r_{a'b'}} \right) \vek{p}_{a}^2
+4\frac{r_{aa'}^2}{r_{a'b'}} (\vek{n}_{aa'}\cdot\vek{p}_{a})^2
\right.
\nonumber\\[1ex]&
\left.
+17\left( \frac{r_{aa'}^2}{r_{a'b'}} + r_{a'b'} \right)
(\vek{n}_{a'b'}\cdot\vek{p}_{a})^2
+ 2 \left( 6 \frac{r_{aa'}^3}{r_{a'b'}^2} + 17 r_{aa'} 
\right) (\vek{n}_{aa'}\cdot\vek{p}_{a}) (\vek{n}_{a'b'}\cdot\vek{p}_{a})
\right]
\label{eq:Rstrich}
\end{align}
and
\begin{align}
R'' \left(\vek{x}_a,t\right):=\,
&\frac{1}{105} \left(\frac{1}{16\pi}\right)^2
\sum\limits_{a}\sum\limits_{b\ne a}\sum\limits_{a'} \frac{m_a m_b}{m_{a'}}
\left[\, \left( 5 \frac{r_{aa'}^2}{r_{ab}}
+ 2 \frac{r_{aa'}^2r_{ba'}^2}{r_{ab}^3}
- 2 \frac{r_{aa'}^4}{r_{ab}^3} \right) \vek{p}_{a'}^2
-17 \left( \frac{r_{aa'}^2}{r_{ab}} + r_{ab} \right)
(\vek{n}_{ab}\cdot\vek{p}_{a'})^2
\right.
\nonumber\\[1ex]&
\left.
- 4 \frac{r_{aa'}^2}{r_{ab}} (\vek{n}_{aa'}\cdot\vek{p}_{a'})^2
\fbox{+}\, 2 \left( \frac{6r_{aa'}^3}{r_{ab}^2} + 17 r_{aa'} \right)
(\vek{n}_{ab}\cdot\vek{p}_{a'}) (\vek{n}_{aa'}\cdot\vek{p}_{a'}) \,\right]
\nonumber\\[1ex]
&+ \frac{1}{210} \left(\frac{1}{16\pi}\right)^3
\sum\limits_{a}\sum\limits_{b\ne a}\sum\limits_{a'}\sum\limits_{b'\ne a'}
m_a m_b m_{a'} m_{b'} \left[\,
2 \frac{r_{aa'}^2}{r_{ab}r_{a'b'}^3}\left(r_{aa'}^2 - r_{ab'}^2\right)
+ 2 \frac{r_{aa'}^2}{\fbox{$r_{ab}^3$}\, r_{a'b'}}\left(r_{aa'}^2 - r_{ba'}^2
\right)
\right.
\nonumber\\[1ex]&
+ 4 \frac{r_{ab}r_{aa'}^2}{r_{a'b'}^3}
- 5 \frac{r_{aa'}^2}{r_{ab}r_{a'b'}}
- 2 \left( \frac{r_{ab}^3}{r_{a'b'}^3} + \frac{r_{ab}}{r_{a'b'}}\right)
+17 \left( \frac{r_{ab}}{r_{a'b'}} + \frac{r_{a'b'}}{r_{ab}}
+ \frac{r_{aa'}^2}{r_{ab}\,\fbox{$r_{a'b'}$}} \right)
(\vek{n}_{ab}\cdot\vek{n}_{a'b'})^2
\nonumber\\[1ex]
&\left. -4 \frac{r_{ab}r_{aa'}r_{bb'}}{r_{a'b'}^3}
(\vek{n}_{aa'}\cdot\vek{n}_{bb'})
+ 6 \frac{r_{aa'}^4}{r_{ab}^2 r_{a'b'}^2}
(\vek{n}_{ab}\cdot\vek{n}_{a'b'}) 
+34r_{aa'}^2\left(\frac{1}{r_{ab}^2}+\frac{1}{r_{a'b'}^2}\right)
(\vek{n}_{ab}\cdot\vek{n}_{a'b'}) \,\right]
\,.
\label{eq:Rbis}
\end{align}
\end{subequations}
\end{widetext}

Now, we are in a position to reduce the $N$-body $3.5$PN Hamiltonian
\eqref{eq:35res} to $N=2$. The result will serve as a base
for the derivation of the gravitational energy loss of a compact binary
moving in general orbits, like in Ref.~\cite{1997Jaranowski}, and
provide a reference to check the equations of motion that will be
calculated afterward from this two-body Hamiltonian.

\section{Energy loss of a two-body system}

Let us denote by $\widetilde{H}_{\le{\rm 3.5PN}}$ the Hamiltonian resulting
from $H_{\le{\rm 3.5PN}}$, Eq.~\eqref{eq:ha} after writing the field part as a
function of $\vek{x}_a, \vek{p}_a, \dot{\vek{x}}_a, \dot{\vek{p}}_a$ and
subtracting its $2.5$PN and $3.5$PN contribution $H_{\rm 2.5PN}^{\rm field}$
and $H_{\rm 3.5PN}^{\rm field}$ from $H_{\le{\rm 3.5PN}}^{\rm field}$. By
definition, we have
\begin{align}
\widetilde{H}_{\le{\rm 3.5PN}}:=\,
&H_{\le{\rm 3.5PN}}^{\rm mat}
+H_{\le{\rm 3.5PN}}^{\rm field}
-H_{\rm 2.5PN}^{\rm field}
-H_{\rm 3.5PN}^{\rm field}
\nonumber \\
&+H_{\le{\rm 3.5PN}}^{\rm int}
\,,
\end{align}
which can be split into a conservative and a dissipative part
\begin{align}
\label{eq:HtildeKonsDiss}
\widetilde{H}_{\le{\rm 3.5PN}}\left(\vek{x}_a,\vek{p}_a,\dot{\vek{x}}_a,
\dot{\vek{p}}_a,t\right)=\,
&H_{\le{\rm 3PN}}^{\rm cons}\left(\vek{x}_a,\vek{p}_a,\dot{\vek{x}}_a,
\dot{\vek{p}}_a \right)
\nonumber \\
&+H_{\le{\rm 3.5PN}}^{\rm diss}\left(\vek{x}_a,\vek{p}_a,t\right)
\,,
\end{align}
where
\begin{subequations}
\begin{align}
H_{\le{\rm 3PN}}^{\rm cons}:=\,
&H_{{\rm N}}^{\rm mat}
+H_{{\rm 1PN}}^{\rm mat}
+\left(
H_{{\rm 2PN}}^{\rm mat}
+H_{{\rm 2PN}}^{\rm field} 
+H_{{\rm 2PN}}^{\rm int}
\right)
\nonumber \\
&+\left(
H_{{\rm 3PN}}^{\rm mat}
+H_{{\rm 3PN}}^{\rm field} 
+H_{{\rm 3PN}}^{\rm int} 
\right)
\,,
\label{eq:HKonsDef}
\\[1mm]
H_{\le{\rm 3.5PN}}^{\rm diss}:=\,
&H_{\rm 2.5PN}^{\rm int}
+H_{\rm 3.5PN}^{\rm int}
\,.
\label{eq:dissdef}
\end{align}
\end{subequations}
The conservative Hamilton function corresponding to the
expression~\eqref{eq:HKonsDef}
is given in Refs.~\cite{JaraSchafer1998PRD57, JaranwoskiSchaefer2000H3PN}.
With the help of the coordinate transformation described in
Ref.~\cite{DamourJaranowskiSchaefer2000}, it is possible to eliminate the
dependence on $\dot{\vek{x}}_a$ and $\dot{\vek{p}}_a$.
The difference between
$H_{\le{\rm 3PN}}^{\rm cons}$ and $H_{\le{\rm 3.5PN}}^{\rm diss}$ in
Eq.~\eqref{eq:HtildeKonsDiss} relates to their behaviors with respect to time
reversal: $H_{\le{\rm 3PN}}^{\rm cons}$ is symmetric whereas
$H_{\le{\rm 3.5PN}}^{\rm diss}$ is antisymmetric.
 
The total time derivative of $\widetilde{H}_{\le{\rm 3.5PN}}$ is equal to its
partial time derivative, the other terms canceling by virtue of the equations
of motion. Only the dissipative part of $\widetilde{H}_{\le{\rm 3.5PN}}$
depends explicitly on time, through the substitution of the
transverse-traceless variables that have been replaced by their actual
values. We get therefore
\begin{align}
\label{eq:timeder}
&\frac{d}{dt}\widetilde{H}_{\le{\rm 3.5PN}}\left(\vek{x}_a, \vek{p}_a,
\dot{\vek{x}}_a, \dot{\vek{p}}_a, t\right)
\nonumber \\
&\quad\quad\quad\quad\quad
=\frac{\pa}{\pa t} \widetilde{H}_{\le{\rm 3.5PN}}\left(\vek{x}_a,
\vek{p}_a, \dot{\vek{x}}_a, \dot{\vek{p}}_a, t\right)
\nonumber \\
&\quad\quad\quad\quad\quad
=\frac{\pa}{\pa t} H_{\le{\rm 3.5PN}}^{\rm diss}\left(\vek{x}_a,
\vek{p}_a, t \right)
\,.
\end{align}
The instantaneous energy loss of the matter system due to the gravitational
wave emission is defined as
\be
{\cal L}_{\le{\rm 3.5PN}}\left(t\right):=
-\frac{d}{dt}\widetilde{H}_{\le{\rm 3.5PN}}\left(\vek{x}_a,\vek{p}_a,
\dot{\vek{x}}_a,\dot{\vek{p}}_a,t\right)\,.
\label{eq:lumdef}
\ee

Out task is now to compute ${\cal L}_{\le{\rm 3.5PN}}$ for a system of two
point-like bodies with masses $m_1$ and $m_2$.\footnote{All algebraic
computations are performed using the commercial software
Mathematica~\cite{Mathematica}.}
We start by rewriting the right-hand side of
Eq.~\eqref{eq:lumdef} [by means of Eqs.~\eqref{eq:dissdef} and
\eqref{eq:timeder}] in the form
\be
{\cal L}_{\le{\rm 3.5PN}}=-\frac{\pa}{\pa t}
\left( H_{\rm 2.5PN}^{\rm int} + H_{\rm 3.5PN}^{\rm int} \right)\,.
\label{eq:loss1}
\ee
Next, we insert the formulas~\eqref{eq:25res} and \eqref{eq:35res} into
Eq.~\eqref{eq:loss1}, and specialize all sums of
Eqs.~\eqref{eq:chidef} and \eqref{H35_is_parametrized_by_the_functions}
for $N=2$. Notice that the triple sum
$\sum\limits_a\sum\limits_{b \ne a}\sum\limits_{c\ne a,b}$ disappears
in relations~\eqref{eq:P1res} and
\eqref{eq:P2res}. For clarity, we also reintroduce the Newtonian gravitational
constant $G$ and the speed of light $c$.

In differentiating the Hamiltonians~\eqref{eq:25res} and \eqref{eq:35res} with
respect to time, we replace the time derivatives of the primed
coordinates and momenta according to the $1$PN equations of motion:
\begin{widetext}
\begin{subequations}
\label{eq:p1strichUNDp2strichNundPN}
\begin{align}
\dot{\vek{x}}_{1'}= &\, \frac{\vek{p}_{1'}}{m_{1'}}-\frac{1}{2 c^2 m_{1'}^3} 
\vek{p}_{1'}^2 \vek{p}_{1'}
-\frac{G}{2 c^2\,r_{1'2'}}
\bigg[ 6 \frac{m_{2'}}{m_{1'}}\vek{p}_{1'} - 7 \vek{p}_{2'}
- \left(\vek{n}_{1'2'}\cdot\vek{p}_{2'}\right) \vek{n}_{1'2'} \bigg]\,,
\label{eq:x1strichNundPN}
\\[1mm]
\dot{\vek{x}}_{2'} =&\, \frac{\vek{p}_{2'}}{m_{2'}}-\frac{1}{2 c^2 m_{2'}^3} 
\vek{p}_{2'}^2 \vek{p}_{2'}
-\frac{G}{2 c^2\,r_{1'2'}}
\bigg[ 6 \frac{m_{1'}}{m_{2'}}\vek{p}_{2'} - 7 \vek{p}_{1'}
- \left(\vek{n}_{1'2'}\cdot\vek{p}_{1'}\right) \vek{n}_{1'2'} \bigg]\,,
\label{eq:x2strichNundPN}
\\[1mm]
\dot{\vek{p}}_{1'} =&\, -\frac{G\,m_{1'}\,m_{2'}}{r_{1'2'}^2} \vek{n}_{1'2'}+
\frac{G^2\,m_{1'}\,m_{2'} (m_{1'}+m_{2'})}{c^2\,r_{1'2'}^3} \vek{n}_{1'2'}
+\frac{G}{2\,c^2\,r_{1'2'}^2}
\Bigg\{
-\left(\vek{n}_{1'2'}\cdot \vek{p}_{2'}\right)\vek{p}_{1'}
-\left(\vek{n}_{1'2'}\cdot\vek{p}_{1'}\right) \vek{p}_{2'}
\nonumber\\[1mm]&
+\bigg[-3 \frac{m_{2'}}{m_{1'}} \vek{p}_{1'}^2 - 
3\frac{m_{1'}}{m_{2'}} \vek{p}_{2'}^2 + 7 \left(\vek{p}_{1'}\cdot
\vek{p}_{2'}\right)
+3 \left(\vek{n}_{1'2'}\cdot\vek{p}_{1'}\right) 
\left(\vek{n}_{1'2'}\cdot\vek{p}_{2'}\right) \bigg] \vek{n}_{1'2'}
\Bigg\} \,, 
\label{eq:p1strichNundPN}
\\[1mm]
\dot{\vek{p}}_{2'} =& \,\frac{G\,m_{1'}\,m_{2'}}{r_{1'2'}^2} \vek{n}_{1'2'}-
\frac{G^2\,m_{1'}\,m_{2'} (m_{1'}+m_{2'})}{c^2\,r_{1'2'}^3} \vek{n}_{1'2'}
-\frac{G}{2\,c^2\,r_{1'2'}^2}
\Bigg\{
-\left(\vek{n}_{1'2'}\cdot
\vek{p}_{2'}\right)\vek{p}_{1'}-\left(\vek{n}_{1'2'}\cdot\vek{p}_{1'}\right)
\vek{p}_{2'}
\nonumber\\[1mm]&
+\bigg[-3 \frac{m_{2'}}{m_{1'}} \vek{p}_{1'}^2 - 
3\frac{m_{1'}}{m_{2'}} \vek{p}_{2'}^2 + 7 \left(\vek{p}_{1'}\cdot
\vek{p}_{2'}\right)
+3 \left(\vek{n}_{1'2'}\cdot\vek{p}_{1'}\right) 
\left(\vek{n}_{1'2'}\cdot\vek{p}_{2'}\right) \bigg] \vek{n}_{1'2'}
\Bigg\}\,.
\label{eq:p2strichNundPN}
\end{align}
\end{subequations}
The calculation of the third time derivative of $R'$ and $R''$ is particularly
tedious.

The particle momenta $\vek{p}_{a/a'}$ is expressed in terms of the particle
coordinate velocities $\vek{v}_{a/a'}$ up to the $1$PN order [see, e.g.,
Eq.~(4.1) of \cite{OOKH74-1220}] by:
\begin{subequations}
\label{eq:changeP1andP2toV}
\begin{align}
\label{eq:changeP1toV_a}
\vek{p}_1 &= m_1 \vek{v}_1 + \frac{1}{2 c^2} m_1 \vek{v}_1^2 \vek{v}_1
+\frac{G\,m_1\,m_2}{2\,c^2\,r_{12}}
\Big[ 6 \vek{v}_1 - 7 \vek{v}_2
- \left(\vek{n}_{12}\cdot\vek{v}_2\right) \vek{n}_{12} \Big]\,,
\\[1mm]
\label{eq:changeP2to_b}
\vek{p}_2 &= m_2 \vek{v}_2 + \frac{1}{2 c^2} m_2 \vek{v}_2^2 \vek{v}_2
+\frac{G\,m_1\,m_2}{2\,c^2\,r_{12}}
\Big[ 6 \vek{v}_2 - 7 \vek{v}_1
- \left(\vek{n}_{12}\cdot\vek{v}_1\right) \vek{n}_{12} \Big]\,.
\end{align}
\end{subequations}
Analogous relations hold for $\vek{p}_{1'}$ and $\vek{p}_{2'}$.

The time derivatives of the primed coordinates $\vek{x}_{1'}$ and
$\vek{x}_{2'}$ coming from the differentiation of
$\widetilde{H}_{\le{\rm 3.5PN}}$ in Eq.~\eqref{eq:lumdef} equate the
primed coordinate
velocities $\vek{v}_{1'}$ and $\vek{v}_{2'}$. We eliminate all accelerations
by making use of the $1$PN equations of motion [see, e.g., Eq.~(1.5) of
\cite{DS88}]:
\begin{subequations}
\begin{align}
 \dot{\vek{v}}_{1'} =&\,
-\frac{G\, m_{2'}}{r_{1'2'}^2}\vek{n}_{1'2'}
+\frac{G\, m_{2'}}{c^2\, r_{1'2'}^2} \Bigg\{
\Big[ 4 \left(\vek{n}_{1'2'}\cdot\vek{v}_{1'}\right)
-3\left(\vek{n}_{1'2'}\cdot\vek{v}_{2'}\right) \Big]
\left(\vek{v}_{1'}-\vek{v}_{2'}\right)
+\bigg[ - \vek{v}_{1'}^2 - 2 \vek{v}_{2'}^2 + 4 \left(\vek{v}_{1'}
\cdot\vek{v}_{2'}\right)
\nonumber\\[1mm]&
\left.
+\frac{3}{2} \left(\vek{n}_{1'2'}\cdot\vek{v}_{2'}\right)^2
+\frac{G}{r_{1'2'}} \left(5 m_{1'} + 4 m_{2'}\right)
\right] \vek{n}_{12} \Bigg\}\,,
\\[1mm]
\dot{\vek{v}}_{2'} =&\,
+\frac{G\, m_{1'}}{r_{1'2'}^2}\vek{n}_{1'2'}
+\frac{G\, m_{1'}}{c^2\, r_{1'2'}^2} \Bigg\{
\Big[4 \left(\vek{n}_{1'2'}\cdot\vek{v}_{2'}\right)
-3\left(\vek{n}_{1'2'}\cdot\vek{v}_{1'}\right) \Big]
\left(\vek{v}_{1'}-\vek{v}_{2'}\right)
-\bigg[ - \vek{v}_{2'}^2 - 2 \vek{v}_{1'}^2 + 4 \left(\vek{v}_{1'}
\cdot\vek{v}_{2'}\right)
\nonumber\\[1mm]&
\left.
+\frac{3}{2} \left(\vek{n}_{1'2'}\cdot\vek{v}_{1'}\right)^2
+\frac{G}{r_{1'2'}} \left(5 m_{2'} + 4 m_{1'}\right)
\right] \vek{n}_{1'2'} \Bigg\}\,.
\end{align}
\end{subequations}
\end{widetext}

At this stage, we can identify positions and velocities of particles
inside and outside the transverse-traceless variables
[\emph{i.e.} the primed and unprimed quantities]. As the limit
\mbox{$\vek{x}_{1'} \rightarrow \vek{x}_{1}$},
\mbox{$\vek{x}_{2'} \rightarrow \vek{x}_{2}$}
is singular, we shall resort to the Hadamard regularization procedure
described in Appendix A [for a more detailed investigation, see
paper~\cite{HadamardRegBlanchetFaye} and the references therein].
The computations are extremely long, but straightforward.

To write the final formula for the energy loss in a more compact way, we
introduce the total mass of the system $M:=m_1+m_2$, the reduced mass
$\mu:=m_1\,m_2/M$, as well as the ratio parameter $\nu:=\mu/M$. The
individual masses $m_1$ and $m_2$ are given in terms of $\mu$ and $\nu$ by:
\begin{subequations}
\label{eq:RelMasses}
\begin{align}
\label{eq:RelMasses_a}
m_1 &= \frac{\mu}{2\nu} \Big(1 + \sqrt{1 - 4\nu}\Big)
\,, \\
\label{eq:RelMasses_b}
m_2 &= \frac{\mu}{2\nu} \Big(1 - \sqrt{1 - 4\nu}\Big)
\,,
\end{align}
\end{subequations}
assuming $m_1\ge m_2$. We also express the velocities of the bodies in the
center-of-mass frame as a function of their relative velocity
\mbox{$\vek{v}=\vek{v_{12}}:= \vek{v}_1 - \vek{v}_2$} and of their separation
\mbox{$\vek{r}=\vek{r_{12}}:= \vek{x}_1 -\vek{x}_2=\vek{r}_2 -\vek{r}_1$} with
\mbox{$\vek{v}=\dot{\vek{r}}$}. This is achieved by means of the relations
linking the positions of the individual particles to $\vek{r}$ at the $1$PN
order [see, e.g., Eq.\ (2.4) of \cite{DD85}]:\footnote{At the $1$PN
order the Eqs.~\eqref{eq:x1x2r12} in  ADM gauge and in harmonic gauge
are identical.}
\begin{subequations}
\label{eq:x1x2r12}
\begin{align}
\label{eq:x1r12_a}
\vek{x}_1 &= \left[ \frac{\mu}{m_1} + \frac{\mu(m_1-m_2)}{2 c^2 M^2}
\left( \vek{v}^2 - \frac{G M}{r} \right) \right] \vek{r}
\,,\\
\label{eq:x2r12_b}
\vek{x}_2 &= \left[-\frac{\mu}{m_2} + \frac{\mu(m_1-m_2)}{2 c^2 M^2}
\left( \vek{v}^2 - \frac{G M}{r} \right) \right] \vek{r}
\,,
\end{align}
\end{subequations}
where $r=|\vek{r}|$. By differentiating Eqs.~\eqref{eq:x1x2r12} with
respect to time and eliminating the temporal derivatives
with the aid of the Newtonian equation of motion
\be
\dot{\vek{v}}=-\frac{G M}{r^2} \vek{n}
\,,
\ee
where $\vek{n}:= \vek{r}/r$, we are led to:
\begin{subequations}
\begin{multline}
\vek{v}_1 =\frac{2 \nu \vek{v}}{1 + \sqrt{1 - 4\nu}} + \frac{\sqrt{1 -
4\nu}}{2 c^2}
\times
\\
\times
\left[ \nu \vek{v}^2 \vek{v} - \frac{G \mu}{r} \Big(
\vek{v} + \left(\vek{n}\cdot\vek{v}\right)\vek{n} \Big) \right]
\label{eq:v1v12}
\end{multline}
\begin{multline}
\vek{v}_2 =\frac{-2 \nu \vek{v}}{1 - \sqrt{1 - 4\nu}} + \frac{\sqrt{1 -
4\nu}}{2 c^2}
\times
\\
\times
\left[ \nu \vek{v}^2 \vek{v} - \frac{G \mu}{r} \Big( \vek{v}
+\left(\vek{n}\cdot\vek{v}\right)\vek{n} \Big) \right] \,.
\label{eq:v2v12}
\end{multline}
\end{subequations}

Finally, we arrive at the instantaneous energy loss~\cite{CK2002}:
\begin{widetext}
\begin{align}
{\cal L}_{\le{\rm 3.5PN}}^{\text{(1)}}=
&\mbox{} \frac{4}{15} \frac{G^2 M^3 \nu^2}{c^5 r^3}
\Bigg\{ 2 \frac{G^2 M^2}{r^2}
+ \Big[ 11 \vek{v}^2 - 9 \left(\vek{n}\cdot\vek{v}\right)^2 \Big]
\frac{G M}{r}
+ \Big[ 11 \vek{v}^4 - 60 \left(\vek{n}\cdot\vek{v}\right)^2 \vek{v}^2
+ 45 \left(\vek{n}\cdot\vek{v}\right)^4 \Big] \Bigg\}
\nonumber\\ 
&+ \frac{1}{105} \frac{G^2 M^3 \nu^2}{c^7 r^3}
\Bigg\{ (5 -104\nu) \frac{G^3 M^3}{r^3}
+\Big[(-1615+776\nu)\vek{v}^2
+(2733+92\nu)\left(\vek{n}\cdot\vek{v}\right)^2
\Big]
\frac{G^2 M^2}{r^2}
\nonumber\\ 
&+\Big[-(1308+1372\nu)\vek{v}^4+4(4158+1577\nu)
\left(\vek{n}\cdot\vek{v}
\right)^2 \vek{v}^2
-3(5174+1496\nu)\left(\vek{n}\cdot\vek{v}\right)^4\Big]\frac{G M}{r}
\nonumber\\ 
&+\Big[-(164+1064\nu)\vek{v}^6+3(716+1562\nu)
\left(\vek{n}\cdot\vek{v}
\right)^2 \vek{v}^4
-15 (324 + 56\nu) \left(\vek{n}\cdot\vek{v}\right)^4 \vek{v}^2
\nonumber\\ 
&+105 (28 - 22 \nu) \left(\vek{n}\cdot\vek{v}\right)^6 \Big] \Bigg\}\,.
\label{eq:loss2}
\end{align}
\end{widetext}

The comparison of the expression above for the
instantaneous gravitational energy loss with the known instantaneous far-zone
flux [see Eq.~(51) of \cite{WW76teil1} corrected by an erratum and equation
below Eq.~(3.40) of \cite{BS89}] shows that they are not either identical.
This is nothing to worry about since the form of the instantaneous
gravitational energy loss in an isolated system is as well \emph{coordinate}
as \emph{representation} dependent.

We also observe that the expression~\eqref{eq:loss2} is different
from formula~(73) [in the work of Jaranowski and
Sch\"afer~\cite{1997Jaranowski}] at the $3.5$PN order.
However, the difference may be expressed as a total time derivative.
To verify that, we computed
$\langle {\cal L}_{\le{\rm 3.5PN}}^{\text{(1)}} \rangle$ and
$\langle {\cal L}_{\le{\rm 3.5PN}}^{\text{JS}} \rangle$, the orbital
average of
${\cal L}_{\le{\rm 3.5PN}}^{\text{(1)}}$ and
${\cal L}_{\le{\rm 3.5PN}}^{\text{JS}}$ respectively, using the
prescription given in Ref.~\cite{BS89}, for quasi-elliptical orbital
motion and found
\mbox{$\langle{\cal L}_{\le{\rm 3.5PN}}^{\rm JS}\rangle =
\langle {\cal L}_{\le{\rm 3.5PN}}^{\text{(1)}} \rangle$}.
So, we recover Eqs.~(4.20) and (4.21) of Ref.~\cite{BS89}.
The important point is that all averaged losses coincide.

In the
following, we shall make the assumption that the binary has entered the
inspiraling phase, so that its relative motion is quasi-circular. The
investigation appears to be simpler in this case, since all terms involving
the factor $\left(\vek{n}\cdot\vek{v}\right)$ vanish. The averaging is
performed by expressing ${\cal L}_{\le{\rm 3.5PN}}^{\text{(1)}}$
as a function of $E$.
According to equation~(3.35) in Ref.~\cite{BS89}, we have for the
energy (reduced by the factor $\mu$)
\begin{align}
\label{eq:Energieformel335}
E=\,&\frac{1}{2}\vek{v}^2-\frac{G M}{r}
+\frac{3}{8 c^2}(1-3 \nu)\vek{v}^4
\nonumber\\
&+\frac{G M}{2 r c^2}\left[(3+\nu) \vek{v}^2+\nu
\left(\vek{n}\cdot\vek{v} \right)^2+\frac{G M}{r}\right]
\,.
\end{align}
The right-hand side depends on the relative velocity $\vek{v}$ and the
relative distance $r$, whereas the symbol $E$ in the left-hand side may be
viewed as an adiabatic parameter. By solving Eq.~\eqref{eq:Energieformel335}
for $\vek{v}^2$, we find
\begin{align}
\label{eq:VVausEnergieformel335}
\vek{v}^2=\,
&2 E+\frac{2 G M}{r}
+\frac{1}{c^2}\bigg[E^2(-3+9 \nu)
\nonumber \\
&+\frac{G M}{r} E (-12+16 \nu)
+\frac{G^2 M^2}{r^2}(-10+7 \nu)\bigg]
\,.
\end{align}
Next, we substitute Eq.~(\ref{eq:VVausEnergieformel335}) to $\vek{v}^2$ in the
instantaneous energy loss~(\ref{eq:loss2}), and eliminate the
relative distance $r$ in favor of the reduced energy with the aid of Eq.~(43)
in Ref.~\cite{Wex1993teil1}:
\begin{align}
r=-\frac{G M}{2 E}\left[1+\frac{1}{2}(7-\nu)\frac{E}{c^2}\right]
\,.
\end{align}
As a consequence, the time-averaged energy loss specialized to the case of
circular orbits (c.o.) reads
\begin{align}
\label{eq:zeitlgemttEloss420Version2}
\Big\langle{\cal L}_{\le{\rm 3.5PN}}\Big\rangle_{t}^{\rm c.o.}=
\frac{1024}{5}\frac{\nu^2 (- E)^5}{c^5 G}
\left[
1+\frac{(-E)}{c^2}\left(\frac{13}{168}- 5 \nu\right)
\right]
\,.
\end{align}
We recover formula~(4.20) in Ref.~\cite{BS89} with $e_{\rm R}=0$
(due to the fact that the orbits are quasi-circular). This is in
agreement with Eq.~(81) of Ref.~\cite{WW76teil1}.
Processing Eq.~\eqref{eq:loss3} or Eq.~(73) in
Ref.~\cite{1997Jaranowski} in a similar way ends in the same result.

The instantaneous gravitational energy loss ${\cal L}_{\le{\rm 3.5PN}}$ given
in Eq.~\eqref{eq:loss2} will be taken as reference to check the equations of
motion that will be determined in the next section.

\section{Hamiltonian equations of motion}

In this section we compute the reaction part of the equations of motion
\be
\dot {\vek p}_a =
-\frac{\pa H^{\rm int}_{\le{\rm 3.5PN}}}{\pa \vek{x}_a}
\quad \quad \text{and}\quad \quad
\dot {\vek x}_a =
\frac{\pa H^{\rm int}_{\le{\rm 3.5PN}}}{\pa \vek{p}_a}
\,,
\ee
in a fully explicit form in terms of $\vek{x}_1, \vek{x}_2,\vek{p}_1$
and $\vek{p}_2$. For this goal, we have to differentiate the
Hamiltonians~\eqref{eq:25res} and \eqref{eq:35res} with respect to the
particle coordinates $\vek{x}_a$ and the particle momenta $\vek{p}_a$ with
$a=1,2$ and apply the Hadamard regularization procedure described in
Appendix A. We find \cite{CK2002}:
\begin{widetext}
\begin{subequations}
\label{eq:x1punktUNDx2punktResultat}
\begin{align}
\dot{\vek{x}}_1 =&\, \frac{1}{c^5}\Bigg\{
\frac{G^2}{r^2}\bigg\{
\left[-\frac{24}{5}\left(\vek{n}\cdot\vek{p}_1 \right)\left(\vek{n}
\cdot \vek{p}_2 \right)+\frac{m_2}{m_1}\left(
\frac{24}{5}\left(\vek{n} \cdot \vek{p}_1 \right)^2 
-\frac{16}{5}\vek{p}_1^2\right)+\frac{16}{5}\left(\vek{p}_1\cdot\vek{p}_2 
\right)\right]\vek{n}
+\left[-\frac{8 m_2}{3 m_1 }\left(\vek{n} \cdot \vek{p}_1 \right)
\right.
\nonumber\\[1mm]&
\left.
-\frac{8}{15}\left(\vek{n}\cdot\vek{p}_2\right)\right]\vek{p}_1
+\frac{16}{5}\left(\vek{n}\cdot\vek{p}_1\right)\vek{p}_2
\bigg\}
\Bigg\}
+\frac{1}{c^7}\Bigg\{ 
\frac{G^2}{r^2}\bigg\{\left[\frac{m_2}{m_1^3}\left(\frac{36}{7}\left(\vek{n} 
\cdot \vek{p}_1 \right)^4 
-\frac{60}{7}\left(\vek{n}\cdot\vek{p}_1\right)^2\vek{p}_1^2
+\frac{4}{35}\vek{p}_1^4\right)\right.
\nonumber\\[1mm]
&+\frac{1}{m_1^2}\left(-\frac{148}{7}\left(\vek{n}\cdot\vek{p}_1\right)^3
\left(\vek{n} \cdot \vek{p}_2 \right) 
+\frac{748}{35}\left(\vek{n} \cdot \vek{p}_1 \right)\left(\vek{n} \cdot 
\vek{p}_2 \right)\vek{p}_1^2
+\frac{12}{7}\left(\vek{n} \cdot \vek{p}_1 \right)^2\left(\vek{p}_1 \cdot 
\vek{p}_2 \right)
+\frac{172}{35}\vek{p}_1^2 \left(\vek{p}_1 \cdot \vek{p}_2 \right)\right)
\nonumber\\[1mm]
&+\frac{1}{m_1 m_2}\left(\frac{104}{7}\left(\vek{n} \cdot \vek{p}_1 \right)^2
\left(\vek{n} \cdot \vek{p}_2 \right)^2 
-\frac{324}{35}\left(\vek{n} \cdot \vek{p}_2 \right)^2\vek{p}_1^2 \right.
-\frac{432}{35}\left(\vek{n} \cdot \vek{p}_1 \right)\left(\vek{n} \cdot
\vek{p}_2 \right)\left(\vek{p}_1 \cdot \vek{p}_2 \right)
-\frac{136}{35}\left(\vek{p}_1 \cdot \vek{p}_2 \right)^2
\nonumber\\[1mm]&
\left. 
+\frac{132}{35}\left(\vek{n} \cdot \vek{p}_1 \right) ^2\vek{p}_2^2
-\frac{172}{105}\vek{p}_1^2\vek{p}_2^2 \right)
+\frac{1}{m_2^2}\left(\frac{8}{7}\left(\vek{n}\cdot\vek{p}_1 \right)\,
\left(\vek{n} \cdot \vek{p}_2 \right)^3 
+\frac{292}{35}\left(\vek{n} \cdot \vek{p}_2 \right)^2\,\left(\vek{p}_1 \cdot 
\vek{p}_2 \right) \right.
\nonumber\\[1mm]
&\left.\left.
-\frac{176}{35}\left(\vek{n} \cdot \vek{p}_1 \right)\left(\vek{n}\cdot 
\vek{p}_2 \right)\vek{p}_2^2
+\frac{52}{105}\left(\vek{p}_1\cdot\vek{p}_2\right)\vek{p}_2^2\right)
\right]\vek{n}
+\left[
\frac{m_2}{m_1^3}\left(-\frac{324}{35}\left(\vek{n} \cdot \vek{p}_1 
\right)^3
+\frac{1672}{105}\left(\vek{n}\cdot\vek{p}_1 \right)\vek{p}_1^2\right)\right.
\nonumber\\[1mm]
&+\frac{1}{m_1^2}\left(\frac{64}{7}\left(\vek{n} \cdot \vek{p}_1 \right)^2
\left(\vek{n} \cdot \vek{p}_2 \right)
-\frac{1136}{105}\left(\vek{n} \cdot \vek{p}_2 \right)\vek{p}_1^2
-\frac{184}{35}\left(\vek{n} \cdot \vek{p}_1 \right)\left(\vek{p}_1 \cdot 
\vek{p}_2 \right)\right)
\nonumber\\[1mm]
&+\frac{1}{m_1 m_2}\left(\frac{76}{7}\left(\vek{n} \cdot \vek{p}_1 \right)
\left(\vek{n} \cdot \vek{p}_2 \right)^2 
+\frac{376}{105}\left(\vek{n} \cdot \vek{p}_2 \right)\left(\vek{p}_1 \cdot 
\vek{p}_2 \right)
-\frac{992}{105}\left(\vek{n} \cdot \vek{p}_1 \right)\vek{p}_2^2\right)
\nonumber\\[1mm]
&\left.
+\frac{1}{m_2^2}\left(
-\frac{208}{35}\left(\vek{n} \cdot \vek{p}_2 \right)^3
+\frac{112}{15}\left(\vek{n}\cdot\vek{p}_2\right)\vek{p}_2^2\right)
\right]
\vek{p}_1
+\left[\frac{1}{m_1^2}\left(\frac{76}{5}\left(\vek{n} \cdot \vek{p}_1\right)^3 
-\frac{108}{5}\left(\vek{n} \cdot \vek{p}_1 \right)\vek{p}_1^2\right)\right.
\nonumber\\[1mm]
&+\frac{1}{m_1 m_2}\left(-\frac{992}{35}\left(\vek{n} \cdot \vek{p}_1 
\right)^2\left(\vek{n} \cdot \vek{p}_2 \right)
+\frac{664}{35}\left(\vek{n} \cdot \vek{p}_2 \right)\vek{p}_1^2 
+\frac{400}{21}\left(\vek{n} \cdot \vek{p}_1 \right)\left(\vek{p}_1 \cdot 
\vek{p}_2 \right)\right)
\nonumber\\[1mm]
&\left.
+\frac{1}{m_2^2}\left(\frac{292}{35}\left(\vek{n} \cdot \vek{p}_1 \right)
\left(\vek{n} \cdot \vek{p}_2 \right)^2 
-\frac{1928}{105}\left(\vek{n} \cdot \vek{p}_2 \right)\left(\vek{p}_1 \cdot 
\vek{p}_2 \right) 
+\frac{52}{105}\left(\vek{n} \cdot \vek{p}_1 \right)\vek{p}_2^2\right)
\right]
\vek{p}_2
\bigg\}
\nonumber\\[1mm]
&+\frac{G^3}{r^3}\bigg\{\left[\frac{m_2^2}{m_1}\left(
-\frac{7862}{105}\left(\vek{n} \cdot \vek{p}_1 \right)^2
+\frac{4129}{105}\vek{p}_1^2\right)
\right. 
+m_2\left(-\frac{248}{5}\left(\vek{n} \cdot \vek{p}_1 \right)^2 
+\frac{3756}{35}\left(\vek{n}\cdot\vek{p}_1 \right)\left(\vek{n} \cdot 
\vek{p}_2 \right)
\right.
\nonumber \\[1mm]
&\left. +\frac{852}{35}\vek{p}_1^2
-\frac{2106}{35}\left(\vek{p}_1 \cdot \vek{p}_2 \right)\right)
\left.
+m_1\left(\frac{5032}{105}\left(\vek{n} \cdot \vek{p}_1 \right)\left(
\vek{n} \cdot \vek{p}_2 \right)
-\frac{158}{5}\left(\vek{n} \cdot \vek{p}_2 \right)^2 
-\frac{484}{21}\left(\vek{p}_1 \cdot \vek{p}_2 \right)
+\frac{97}{5}\vek{p}_2^2\right)
\right]
\vek{n} 
\nonumber\\[1mm]
&+\left[
\frac{617 m_2^2}{15 m_1}\left(\vek{n} \cdot \vek{p}_1 \right)
+m_2\left(\frac{2588}{105}\left(\vek{n} \cdot \vek{p}_1 \right)
-\frac{431}{21}\left(\vek{n} \cdot \vek{p}_2 \right) \right)
-\frac{64}{105} m_1 \left(\vek{n} \cdot \vek{p}_2 \right)
\right]
\vek{p}_1
\nonumber\\[1mm]
&+\left[-\frac{1469}{35} m_2 \left(\vek{n} \cdot \vek{p}_1 \right)
+ m_1 \left(-\frac{484}{21}\left(\vek{n} \cdot \vek{p}_1 \right)
+\frac{103}{5}\left(\vek{n} \cdot \vek{p}_2 \right)\right)
\right]
\vek{p}_2
\bigg\}
+\frac{G^4}{r^4}\Big(m_1^2 m_2^2 + m_1 m_2^3 \Big)\vek{n}
\Bigg\}
\,,
\label{eq:x1punktResultat}
\\[1mm]
\dot{\vek{x}}_2 = &\,(1 \rightleftharpoons 2)
\,.
\label{eq:x2punktResultat}
\end{align}
\end{subequations}
For the time derivative of the particle momenta, we get
\begin{subequations}
\label{eq:p1punktUNDp2punktResultat}
\begin{align}
\dot{\vek{p}}_1 =&\, \frac{1}{c^5} 
\Bigg\{
\frac{G^3}{r^4} 
\bigg\{
\left[
\frac{56}{15} m_1 m_2^2 \left(\vek{n} \cdot \vek{p}_1 \right) 
-\frac{56}{15} m_1^2 m_2\left(\vek{n}\cdot\vek{p}_2 \right)
\right]
\vek{n}
-\frac{16}{5} m_1 m_2^2\,\vek{p}_1 
+\frac{16}{5} m_1^2 m_2\,\vek{p}_2
\bigg\}
\Bigg\} 
\nonumber\\[1mm]
&+\frac{1}{c^7}\Bigg\{
\frac{G^2}{r^3}\bigg\{
\left[
\frac{m_2}{m_1^3}\left(-10\left(\vek{n} \cdot \vek{p}_1 \right)^5 
+\frac{116}{7}\left(\vek{n} \cdot \vek{p}_1 \right)^3\vek{p}_1^2 
-\frac{198}{35}\left(\vek{n} \cdot \vek{p}_1 \right)\vek{p}_1^4
\right)
\right.
\nonumber\\[1mm]
&+\frac{1}{m_1^2}\left(30\left(\vek{n} \cdot \vek{p}_1 \right)^4\left(\vek{n} 
\cdot\vek{p}_2\right)
-\frac{180}{7}\left(\vek{n} \cdot \vek{p}_1 \right)^2\left(\vek{n}\cdot 
\vek{p}_2 \right)\vek{p}_1^2
+\frac{18}{7}\left(\vek{n} \cdot \vek{p}_2 \right)\vek{p}_1^4 
-\frac{184}{7}\left(\vek{n} \cdot \vek{p}_1 \right)^3\left(\vek{p}_1 \cdot 
\vek{p}_2 \right)
\right.
\nonumber\\[1mm]
&\left.
+\frac{568}{35}\left(\vek{n}\cdot\vek{p}_1\right)\vek{p}_1^2\left(\vek{p}_1 
\cdot\vek{p}_2\right)
\right)
+\frac{1}{m_1 m_2}
\bigg( - 30 \left(\vek{n}\cdot\vek{p}_1\right)^3 \left(\vek{n}\cdot\vek{p}_2
\right)^2 
+6\left(\vek{n}\cdot\vek{p}_1\right)\left(\vek{n}\cdot\vek{p}_2\right)^2
\vek{p}_1^2
\nonumber\\[1mm]
&+44\left(\vek{n}\cdot\vek{p}_1\right)^2\left(\vek{n}\cdot\vek{p}_2\right)
\left(\vek{p}_1\cdot\vek{p}_2\right)
-\frac{228}{35}\left(\vek{n}\cdot\vek{p}_2\right)\vek{p}_1^2\left(\vek{p}_1 
\cdot\vek{p}_2\right)
-\frac{456}{35}\left(\vek{n}\cdot\vek{p}_1\right)\left(\vek{p}_1\cdot\vek{p}_2
\right)^2
\nonumber\\[1mm]
&\left.
+\frac{30}{7}\left(\vek{n}\cdot\vek{p}_1\right)^3\vek{p}_2^2 
-\frac{38}{35}\left(\vek{n}\cdot\vek{p}_1\right) \vek{p}_1^2 \,\vek{p}_2^2 
\right)
+\frac{1}{m_2^2}\left(
10\left(\vek{n}\cdot\vek{p}_1\right)^2\left(\vek{n} 
\cdot\vek{p}_2\right)^3 
+\frac{22}{7}\left(\vek{n}\cdot\vek{p}_2\right)^3\vek{p}_1^2
\right.
\nonumber \\[1mm]
&\left.-\frac{124}{7}\left(\vek{n}\cdot\vek{p}_1\right)
\left(\vek{n}\cdot\vek{p}_2 
\right)^2\left(\vek{p}_1\cdot\vek{p}_2\right)
\right.
+\frac{208}{35}\left(\vek{n}\cdot\vek{p}_2\right)\left(\vek{p}_1\cdot\vek{p}_2
\right)^2
-\frac{30}{7}\left(\vek{n} \cdot \vek{p}_1 \right)^2\left(\vek{n} \cdot 
\vek{p}_2 \right)\vek{p}_2^2
\nonumber\\[1mm]
&\left.\left.
-2\left(\vek{n}\cdot\vek{p}_2\right)\vek{p}_1^2 \,\vek{p}_2^2 
+\frac{124}{35}\left(\vek{n}\cdot\vek{p}_1\right)\left(\vek{p}_1\cdot\vek{p}_2
\right)\vek{p}_2^2 
\right)
\right] 
\vek{n} 
+\left[\frac{m_2}{m_1^3}\left(-\frac{26}{7}\left(\vek{n}\cdot\vek{p}_1\right)^4
+4\left(\vek{n}\cdot\vek{p}_1\right)^2\vek{p}_1^2
-\frac{62}{105}\vek{p}_1^4 
\right)
\right.
\nonumber\\[1mm]
&+\frac{1}{m_1^2}\left(
\frac{40}{7}\left(\vek{n}\cdot\vek{p}_1\right)^3
\left(\vek{n}\cdot\vek{p}_2\right)
-\frac{16}{5}\left(\vek{n}\cdot\vek{p}_1\right) \left(\vek{n} \cdot 
\vek{p}_2 \right)\vek{p}_1^2 
-\frac{8}{5}\left(\vek{n}\cdot\vek{p}_1\right)^2\left(\vek{p}_1\cdot\vek{p}_2
\right)
+\frac{32}{105}\vek{p}_1^2 \left(\vek{p}_1\cdot\vek{p}_2\right)
\right)
\nonumber\\[1mm]
&+\frac{1}{m_1 m_2}\left(
-\frac{2}{7}\left(\vek{n}\cdot\vek{p}_1\right)^2
\left(\vek{n}\cdot\vek{p}_2\right)^2
+\frac{2}{7}\left(\vek{n}\cdot\vek{p}_2 \right)^2\vek{p}_1^2
-\frac{8}{5}\left(\vek{n}\cdot\vek{p}_1 \right)\left(\vek{n}\cdot\vek{p}_2 
\right)\left(\vek{p}_1\cdot\vek{p}_2\right)
\right.
\nonumber\\[1mm]
&\left.
+\frac{16}{21}\left(\vek{p}_1\cdot\vek{p}_2\right)^2 
+\frac{2}{7}\left(\vek{n}\cdot\vek{p}_1\right)^2\vek{p}_2^2
-\frac{22}{105}\vek{p}_1^2 \,\vek{p}_2^2
\right) 
+\frac{1}{m_2^2}\left(
-\frac{12}{7}\left(\vek{n}\cdot\vek{p}_1\right)\left(
\vek{n}\cdot\vek{p}_2\right)^3
+\frac{36}{35}\left(\vek{n}\cdot\vek{p}_2\right)^2\left(\vek{p}_1\cdot
\vek{p}_2 \right)
\right.
\nonumber\\[1mm]
&\left.\left.
+\frac{4}{5}\left(\vek{n} \cdot \vek{p}_1 \right)\left(\vek{n}\cdot\vek{p}_2 
\right)\vek{p}_2^2
-\frac{4}{15}\left(\vek{p}_1\cdot\vek{p}_2\right)\vek{p}_2^2\right)
\right]
\vek{p}_1
+\left[ 
\frac{1}{m_1^2}\left(-\frac{6}{7}\left(\vek{n}\cdot\vek{p}_1\right)^4 
+\frac{12}{5}\left(\vek{n}\cdot\vek{p}_1\right)^2\vek{p}_1^2
-\frac{22}{35}\vek{p}_1^4
\right)
\right.
\nonumber\\[1mm]
&+\frac{1}{m_1 m_2}\left(
\frac{12}{7}\left(\vek{n}\cdot\vek{p}_1\right)^3\left(\vek{n}\cdot\vek{p}_2 
\right)
-\frac{12}{35}\left(\vek{n}\cdot\vek{p}_1\right)\left(\vek{n}\cdot\vek{p}_2 
\right)\vek{p}_1^2
-\frac{204}{35}\left(\vek{n}\cdot\vek{p}_1\right)^2\left(\vek{p}_1\cdot
\vek{p}_2 \right)
+\frac{28}{15}\vek{p}_1^2 \left(\vek{p}_1\cdot\vek{p}_2\right)
\right) 
\nonumber\\[1mm]
&+\frac{1}{m_2^2}\left(
-\frac{6}{7}\left(\vek{n}\cdot\vek{p}_1\right)^2\left(
\vek{n}\cdot\vek{p}_2\right)^2
-\frac{66}{35}\left(\vek{n}\cdot\vek{p}_2\right)^2\vek{p}_1^2
+\frac{192}{35}\left(\vek{n}\cdot\vek{p}_1\right)\left(\vek{n}\cdot\vek{p}_2 
\right)\left(\vek{p}_1\cdot\vek{p}_2\right)
-\frac{40}{21}\left(\vek{p}_1\cdot\vek{p}_2\right)^2
\right.
\nonumber\\[1mm]
&\left.\left.
+\frac{6}{35}\left(\vek{n}\cdot\vek{p}_1\right)^2\vek{p}_2^2
+\frac{2}{3}\vek{p}_1^2 \,\vek{p}_2^2
\right)
\right] 
\vek{p}_2
\bigg\}
+\frac{G^3}{r^4}\bigg\{
\left[
\frac{m_2^2}{m_1}\left(-\frac{52}{5}\left(\vek{n}\cdot\vek{p}_1\right)^3
+\frac{1408}{105}\left(\vek{n}\cdot \vek{p}_1 \right) \vek{p}_1^2 \right)
\right.
\nonumber\\[1mm]
&+m_2 \left(\frac{20}{7}\left(\vek{n} \cdot \vek{p}_1 \right)^3
+\frac{84}{5}\left(\vek{n}\cdot\vek{p}_1\right)^2\left(\vek{n}\cdot\vek{p}_2 
\right)
-\frac{124}{105}\left(\vek{n} \cdot \vek{p}_1 \right)\vek{p}_1^2
-\frac{1388}{105}\left(\vek{n}\cdot\vek{p}_2\right)\vek{p}_1^2 
\right.
\nonumber\\[1mm]
&\left.
-\frac{3128}{105}\left(\vek{n}\cdot\vek{p}_1\right)\left(\vek{p}_1\cdot 
\vek{p}_2 \right)
\right)
+m_1 \left(-\frac{20}{7}\left(\vek{n}\cdot\vek{p}_1\right)^2\left(\vek{n} 
\cdot\vek{p}_2\right) 
-\frac{464}{35}\left(\vek{n}\cdot\vek{p}_1\right)\left(\vek{n}\cdot \vek{p}_2
\right)^2
+\frac{68}{105}\left(\vek{n} \cdot \vek{p}_2 \right)\vek{p}_1^2
\right.
\nonumber\\[1mm]
&\left.
+\frac{24}{35}\left(\vek{n}\cdot\vek{p}_1 \right)\left(\vek{p}_1 \cdot 
\vek{p}_2 \right)
+\frac{2624}{105}\left(\vek{n} \cdot \vek{p}_2 \right)\left(\vek{p}_1 \cdot 
\vek{p}_2 \right)
+\frac{1472}{105}\left(\vek{n} \cdot \vek{p}_1 \right)\vek{p}_2^2 
\right)
\nonumber\\[1mm]
&\left.
+\frac{m_1^2}{m_2}\left( 
\frac{48}{7}\left(\vek{n} \cdot \vek{p}_2 \right)^3
-\frac{324}{35}\left(\vek{n} \cdot \vek{p}_2 \right)\vek{p}_2^2 
\right)
\right]
\vek{n}
+\left[
\frac{m_2^2}{m_1}\left(-\frac{62}{21}\left(\vek{n} \cdot \vek{p}_1 \right)^2
+\frac{82}{105}\vek{p}_1^2\right)
\right.
\nonumber\\[1mm]
&+m_2 \left(-\frac{128}{105}\left(\vek{n}\cdot\vek{p}_1\right)^2
+\frac{1664}{105}\left(\vek{n}\cdot\vek{p}_1\right)\left(\vek{n}\cdot\vek{p}_2
\right)
+\frac{8}{35}\vek{p}_1^2
+\frac{188}{105}\left(\vek{p}_1\cdot\vek{p}_2\right) 
\right)
\nonumber\\[1mm]
&\left.
+m_1 \left(\frac{8}{35}\left(\vek{n}\cdot\vek{p}_1\right)\left(\vek{n}\cdot 
\vek{p}_2 \right)
-\frac{58}{15}\left(\vek{n}\cdot\vek{p}_2\right)^2
+\frac{8}{105}\left(\vek{p}_1\cdot\vek{p}_2\right)
-\frac{34}{105}\vek{p}_2^2 
\right)
\right]
\vek{p}_1
\nonumber\\[1mm]
&+\left[ 
m_2 \left(\frac{134}{21}\left(\vek{n}\cdot\vek{p}_1 \right)^2
-\frac{254}{105}\vek{p}_1^2 
\right)
\right.
+m_1\left(\frac{148}{105}\left(\vek{n} \cdot \vek{p}_1 \right)^2
-\frac{2024}{105}\left(\vek{n}\cdot\vek{p}_1\right)\left(\vek{n}\cdot\vek{p}_2
\right)
\right.
\nonumber \\[1mm]
&\left.-\frac{52}{105}\vek{p}_1^2
+\frac{292}{105}\left(\vek{p}_1\cdot\vek{p}_2\right)
\right)
\left.+\frac{m_1^2}{m_2}\left(\frac{30}{7}\left(\vek{n}\cdot\vek{p}_2\right)^2
-\frac{14}{5}\vek{p}_2^2
\right)
\right]
\vek{p}_2
\bigg\}
+\frac{G^4}{r^5}\bigg\{
\left[
-\frac{3574}{105} m_1 m_2^3\left(\vek{n}\cdot\vek{p}_1\right)
\right.
\nonumber \\[1mm]
& \left.+\frac{174}{5}m_1^3\,m_2\left(\vek{n} \cdot \vek{p}_2 \right) 
+m_1^2 m_2^2\left(-\frac{3838}{105}\left(\vek{n} \cdot \vek{p}_1 \right)
+\frac{3782}{105}\left(\vek{n}\cdot\vek{p}_2\right) \right)
\right]
\vek{n}
\nonumber\\[1mm]
&+\left[ 
\frac{692}{21} m_1^2 m_2^2
+\frac{3268}{105} m_1 m_2^3
\right]
\vek{p}_1
+\left[ 
-\frac{1116}{35} m_1^3 m_2 
-\frac{164}{5} m_1^2 m_2^2 
\right]
\vek{p}_2
\bigg\}
\Bigg\}
\,,
\label{eq:p1punktResultat}
\\[1mm]
\label{eq:p2punktResultat}
\dot{\vek{p}}_2 = &\, (1 \rightleftharpoons 2 )
\,.
\end{align}
\end{subequations}

To check the correctness of the expressions for $\dot{\vek{x}}_a$,
$\dot{\vek{p}}_a$, we employ them to compute the instantaneous
energy loss of a binary, at $2.5$PN and $3.5$PN orders by
differentiating $1$PN  accurate conservative Hamiltonian expression
[see, Hamiltonian~(2.5) of Ref.~\cite{DS88} truncated at the
$1$PN level].\footnote{The $2$PN terms do not play any role in the
computation of the $2.5$PN and $3.5$PN energy loss, since they first
couple with the reaction force at the $(2+2.5)$PN = $4.5$PN order.}
The contribution to the Hamiltonian above comes purely from the
matter part and it reads
\begin{align}
\label{eq:HAMILTONfkt-1PN-ADM}
H_{\rm \leq 1PN}^{\rm mat}(\vek{x}_a,\vek{p}_a)=
&\mbox{}\frac{\vek{p}_{1}^2}
{2 m_1} + \frac{\vek{p}_{2}^2}{2 m_2} - \frac{G m_1 m_2}{r} 
-\frac{1}{8 c^2}\left(\frac{\vek{p}_{1}^4}{m_{1}^3}
+ \frac{\vek{p}_{2}^4}{m_{2}^3} \right)
+\frac{G m_1 m_2}{2 r c^2} \left[
-3 \left( \frac{\vek{p}_{1}^2}{m_1^2}
+ \frac{\vek{p}_{2}^2}{m_2^2} \right)
\right.
\nonumber \\
& \left.
+ 7 \frac{(\vek{p}_{1} \cdot \vek{p}_{2})}{m_1 m_2}
+ \frac{(\vek{n} \cdot \vek{p}_{1})(\vek{n}\cdot\vek{p}_{2})}{m_1 m_2}\right] 
+\frac{G^2 m_1 m_2 (m_1+m_2)}{2 r^2 c^2}\,.
\end{align}
Differentiation with respect to time leads to
\be
\label{eq:ddtHmat1PN}
\frac{d}{dt}H_{\rm \leq 1PN}^{\rm mat}=
\frac{\pa H_{\rm \leq 1PN}^{\rm mat}}{\pa\vek{x}_1}\dot{\vek{x}}_1
+\frac{\pa H_{\rm \leq 1PN}^{\rm mat}}{\pa\vek{x}_2}\dot{\vek{x}}_2
+\frac{\pa H_{\rm \leq 1PN}^{\rm mat}}{\pa\vek{p}_1}\dot{\vek{p}}_1
+\frac{\pa H_{\rm \leq 1PN}^{\rm mat}}{\pa\vek{p}_2}\dot{\vek{p}}_2\,.
\ee 
We then replace the time derivatives of $\vek{x}_1$, $\vek{x}_2$, $\vek{p}_1$
and $\vek{p}_2$ occurring in Eq.~(\ref{eq:ddtHmat1PN}) by their
expressions~\eqref{eq:x1punktUNDx2punktResultat} and
\eqref{eq:p1punktUNDp2punktResultat}.
At last, we write the particle momenta in terms of the particle coordinate
velocities with the help of relations~\eqref{eq:changeP1andP2toV}.
We introduce the relative velocity, the relative
position, the total mass and the parameter $\nu$,
and multiply by $-1$ according to definition~\eqref{eq:lumdef}.
In this way, we obtain for the luminosity~\cite{CK2002}:
\begin{align}
{\cal L}_{\le{\rm 3.5PN}}^{\text{(2)}}=
-\frac{d}{dt}H_{\rm \leq 1PN}^{\rm mat}=\,
&\frac{4}{15} \frac{G^2 M^3 \nu^2}{c^5 r^3}
\Bigg\{
\Big[24 \vek{v}^2 -22 \left(\vek{n}\cdot\vek{v}\right)^2 \Big]
\frac{G M}{r}
\Bigg\}
\nonumber\\
&+\frac{1}{105} \frac{G^2 M^3 \nu^2}{c^7 r^3}
\Bigg\{ -105 \frac{G^3 M^3}{r^3}
+\Big[-(4709+384\nu)\vek{v}^2
+(4821+280\nu)\left(\vek{n}\cdot\vek{v}\right)^2
\Big]
\frac{G^2 M^2}{r^2}
\nonumber\\
&+\Big[(914-784\nu)\vek{v}^4 - 2(1313-1894\nu)
\left(\vek{n}\cdot\vek{v}
\right)^2 \vek{v}^2
+ 3(508-1012\nu)\left(\vek{n}\cdot\vek{v}\right)^4\Big]\frac{G M}{r}
\nonumber\\
&+\Big[(62-344\nu)\vek{v}^6
+3(58+284\nu)\left(\vek{n}\cdot\vek{v}
\right)^2 \vek{v}^4
-450(3-4\nu) \left(\vek{n}\cdot\vek{v}\right)^4 \vek{v}^2
\nonumber\\
&+1050(1-2\nu) \left(\vek{n}\cdot\vek{v}
\right)^6 \Big] \Bigg\}\,.
\label{eq:loss3}
\end{align}
\end{widetext}
Though this result differs from the instantaneous energy loss
\eqref{eq:loss2}, one will see by using the method presented in
Ref.~\cite{BS89} for averaging that it leads to the same averaged value, in
agreement with Eqs.~(4.20) and (4.21) of Ref.~\cite{BS89}.

\section{Euler-Lagrangian equations of motions}

The aim of the present section is to compute the PN reactive corrections
to $\ddot{\vek{x}}_a$ employing two slightly different methods. We will
see that the computation of these Euler-Lagrangian equations of motion
does not involve a Lagrangian in either of these
methods. However, we will require PN accurate relations between particle
momenta and its coordinate velocities. We turn to derive these relations
in the next subsection.

\subsection{Legendre transformation}

For the present computation, we require $1$PN, $2.5$PN and $3.5$PN
corrections to the familiar $\vek{p}_a = m_a \vek{v}_a$ relations.
So, we have to invert the
equations~\eqref{eq:x1punktUNDx2punktResultat} giving
$\dot{\vek{x}}_1=f(\vek{x}_1, \vek{x}_2,\vek{p}_1,\vek{p}_2)$ and
$\dot{\vek{x}}_2$ respectively, in order to determine
$\vek{p}_1=f(\vek{x}_1,\vek{x}_2,\dot{\vek{x}}_1,\dot{\vek{x}}_2)=
f(\vek{x}_1,\vek{x}_2,\vek{v}_1,\vek{v}_2)$ and $\vek{p}_2$.
This will provide the link between the particle momenta and
the particle coordinate velocities up to the order $1/c^7$.

As the first step, we add Newtonian and $1$PN contributions to
$\dot{\vek{x}}_a$ to the right-hand side of
Eqs.~\eqref{eq:x1punktUNDx2punktResultat}. These corrections are
obtained from Eqs.~\eqref{eq:x1strichNundPN} and
\eqref{eq:x2strichNundPN} after replacing primed quantities by their
unprimed counterparts. Through the expression presented below refer to
the first body, it should be noted that similar expressions hold true
for the second body. It is the full coupled system that must be
inverted. The time derivative of the position $\vek{x}_1$ has the form
\begin{align}
\label{eq:Ansatzv1INV}
\dot{\vek{x}}_1=\vek{v}_1=\frac{\vek{p}_1}{m_1}+\frac{1}{c^2}A_2
+\frac{1}{c^5}A_5+\frac{1}{c^7}A_7\,,
\end{align}
whereas the structure of $\vek{p}_1$ after inversion is
\begin{align}
\label{eq:Ansatzp1INV}
\vek{p}_1=m_1 \vek{v}_1+\frac{1}{c^2}B_2
+\frac{1}{c^5}B_5+\frac{1}{c^7}B_7\,. 
\end{align}
The coefficients $B_2, B_5$ and $B_7$ in Eq.~\eqref{eq:Ansatzp1INV} are
computed iteratively in terms of $A_2, A_5$ and $A_7$ of by means of
Eq.~\eqref{eq:Ansatzv1INV}. The first-order coefficient $B_2$ is read directly
from Eq.~\eqref{eq:changeP1toV_a}. We get \cite{CK2002}:
\begin{widetext}
\begin{subequations}
\label{eq:p1UNDp2ausInvertierung}
\begin{align}
\vek{p}_1 = &\; m_1 \vek{v}_1
+\frac{1}{c^2}
\Bigg\{
\frac{1}{2} m_1 \vek{v}_1^2 \,\vek{v}_1
+\frac{G m_1 m_2}{2 r}\bigg\{-\left(\vek{n}\cdot\vek{v}_2\right)
\vek{n}+6\vek{v}_1
-7\vek{v}_2\bigg\}
\Bigg\}
+\frac{1}{c^5}
\Bigg\{
\frac{G^2 m_1^2 m_2}{r^2}
\bigg\{
\left[
-\frac{24}{5}\left(\vek{n}\cdot\vek{v}_1\right)^2
\right.
\nonumber\\[1mm]&
\left.
+\frac{24}{5}\left(\vek{n}\cdot\vek{v}_1\right)
\left(\vek{n}\cdot\vek{v}_2 \right)
+\frac{16}{5}\vek{v}_1^2
-\frac{16}{5}\left(\vek{v}_1\cdot\vek{v}_2\right)
\right]
\vek{n}
+\left[
\frac{8}{3}\left(\vek{n}\cdot\vek{v}_1\right)
+\frac{8}{15}\left(\vek{n}\cdot\vek{v}_2\right)
\right]
\vek{v}_1
-\frac{16}{5}
\left(\vek{n}\cdot\vek{v}_1\right)
\vek{v}_2
\bigg\}
\Bigg\}
\nonumber\\[1mm]
&+\frac{1}{c^7}
\Bigg\{
\frac{G^2 m_1^2 m_2}{r^2}
\bigg\{
\left[
-\frac{36}{7}\left(\vek{n}\cdot\vek{v}_1\right)^4
+\frac{148}{7}\left(\vek{n}\cdot\vek{v}_1\right)^3\left(\vek{n}\cdot\vek{v}_2
\right)
\right.
-\frac{104}{7}\left(\vek{n}\cdot\vek{v}_1\right)^2\left(\vek{n}\cdot\vek{v}_2
\right)^2
-\frac{8}{7}\left(\vek{n}\cdot\vek{v}_1\right)\left(\vek{n}\cdot\vek{v}_2
\right)^3
\nonumber\\[1mm]&
+\frac{164}{35}\vek{v}_1^4
+\left(
-\frac{12}{7}\left(\vek{n}\cdot\vek{v}_1\right)^2
+\frac{432}{35}\left(\vek{n}\cdot\vek{v}_1\right)\left(\vek{n}\cdot\vek{v}_2
\right)
-\frac{292}{35}\left(\vek{n}\cdot\vek{v}_2\right)^2\right)\left(\vek{v}_1\cdot
\vek{v}_2\right)
+\frac{136}{35}\left(\vek{v}_1\cdot\vek{v}_2\right)^2
\nonumber\\[1mm]&
+\left(
-\frac{132}{35}\left(\vek{n}\cdot\vek{v}_1\right)^2
+\frac{52}{7}\left(\vek{n}\cdot\vek{v}_1\right)\left(\vek{n}\cdot\vek{v}_2
\right)
-\frac{44}{21}\left(\vek{v}_1\cdot\vek{v}_2\right)\right)\vek{v}_2^2
+\left(
\frac{48}{35}{\left(\vek{n}\cdot\vek{v}_1\right)}^2
-\frac{116}{7}\left(\vek{n}\cdot\vek{v}_1\right)\left(\vek{n}\cdot\vek{v}_2
\right)
\right.
\nonumber\\[1mm]&
\left.
\left.
+\frac{324}{35}{\left(\vek{n}\cdot\vek{v}_2\right)}^2
-\frac{284}{35}\left(\vek{v}_1\cdot\vek{v}_2\right)
+\frac{172}{105}\vek{v}_2^2\right)\vek{v}_1^2
\right] 
\vek{n}
+\left[
\frac{156}{35}\left(\vek{n}\cdot\vek{v}_1\right)^3
-\frac{152}{35}\left(\vek{n}\cdot\vek{v}_1\right)^2\left(\vek{n}\cdot\vek{v}_2
\right)
\right.
\nonumber\\[1mm]&
-\frac{76}{7}\left(\vek{n}\cdot\vek{v}_1\right)\left(\vek{n}\cdot\vek{v}_2
\right)^2
+\frac{208}{35}\left(\vek{n}\cdot\vek{v}_2\right)^3
+\left(-\frac{212}{35}\left(\vek{n}\cdot\vek{v}_1\right)
+\frac{416}{35}\left(\vek{n}\cdot\vek{v}_2\right)\right)\vek{v}_1^2
\nonumber\\[1mm]&
\left.
+\left(
-\frac{8}{7}\left(\vek{n}\cdot\vek{v}_1\right)
-\frac{376}{105}\left(\vek{n}\cdot\vek{v}_2\right)\right)
\left(\vek{v}_1\cdot\vek{v}_2\right)
+\left(
\frac{992}{105}\left(\vek{n}\cdot\vek{v}_1\right)
-\frac{36}{5}\left(\vek{n}\cdot\vek{v}_2\right)\right)
\vek{v}_2^2
\right]\vek{v}_1
\nonumber\\[1mm]&
+\left[
-\frac{76}{5}\left(\vek{n}\cdot\vek{v}_1\right)^3
+\frac{992}{35}\left(\vek{n}\cdot\vek{v}_1\right)^2\left(\vek{n}\cdot\vek{v}_2
\right)
\right.
-\frac{292}{35}\left(\vek{n}\cdot\vek{v}_1\right)\left(\vek{n}\cdot\vek{v}_2
\right)^2
+\left(
\frac{92}{5}\left(\vek{n}\cdot\vek{v}_1\right)
-\frac{664}{35}\left(\vek{n}\cdot\vek{v}_2\right)
\right)
\vek{v}_1^2
\nonumber\\[1mm]&
\left.
+\left(
-\frac{400}{21}\left(\vek{n}\cdot\vek{v}_1\right)
+\frac{1928}{105}\left(\vek{n}\cdot\vek{v}_2\right)
\right)
\left(\vek{v}_1\cdot\vek{v}_2\right)
-\frac{44}{21}\left(\vek{n}\cdot\vek{v}_1\right)\vek{v}_2^2
\right]
\vek{v}_2
\bigg\}
\nonumber\\[1mm]
&+\frac{G^3}{r^3}
\bigg\{
\left[
m_1^3 m_2\left(
\frac{168}{5}{\left(\vek{n}\cdot\vek{v}_1\right)}^2
-\frac{704}{21}\left(\vek{n}\cdot\vek{v}_1\right)\left(\vek{n}\cdot\vek{v}_2
\right)
-\frac{92}{7}\vek{v}_1^2
\right.
+\frac{1412}{105}\left(\vek{v}_1\cdot\vek{v}_2\right)\right)
\nonumber\\[1mm]&
+m_1^2m_2^2\left(
\frac{3326}{105}\left(\vek{n}\cdot\vek{v}_1\right)^2
-\frac{2812}{105}\left(\vek{n}\cdot\vek{v}_1\right)\left(\vek{n}\cdot\vek{v}_2
\right)
-\frac{62}{15}\left(\vek{n}\cdot\vek{v}_2\right)^2
-\frac{221}{21}\vek{v}_1^2
\left.
+\frac{202}{35}\left(\vek{v}_1\cdot\vek{v}_2\right)
+\frac{23}{5}\vek{v}_2^2
\right)
\right]
\vek{n}
\nonumber\\[1mm]&
+\left[
m_1^2 m_2^2\left(
-\frac{257}{15}\left(\vek{n}\cdot\vek{v}_1\right)
+\frac{13}{7}\left(\vek{n}\cdot\vek{v}_2\right)
\right)
+m_1^3 m_2\left(
-\frac{1636}{105}\left(\vek{n}\cdot\vek{v}_1\right)
+\frac{232}{105}\left(\vek{n}\cdot\vek{v}_2\right)
\right)
\right]
\vek{v}_1
\nonumber\\[1mm]&
+\left[
\frac{1412}{105}m_1^3 m_2\left(\vek{n}\cdot\vek{v}_1\right)
+m_1^2 m_2^2\left(
\frac{1607}{105}\left(\vek{n}\cdot\vek{v}_1\right)
-\frac{1}{3}\left(\vek{n}\cdot\vek{v}_2\right)
\right)
\right]
\vek{v}_2
\bigg\}
+\frac{G^4}{r^4}
\bigg\{
\bigg[-m_1^3 m_2^2 -m_1^2 m_2^3\bigg]\vek{n}
\bigg\}
\Bigg\}
\,,
\label{eq:p1ausInvertierung}
\\[1mm]
\vek{p}_2= &\; (1 \rightleftharpoons 2)
\label{eq:p2ausInvertierung}
\,.
\end{align}
\end{subequations}

\subsection{First method to calculate the Euler-Lagrangian equations
of motions}

The calculation of the accelerations makes the connection with the
Euler-Lagrangian approach. It is achieved in a first stage by evaluating the
total time derivatives of the functions
$\dot{\vek{x}}_1=f(\vek{x}_1,\vek{x}_2,\vek{p}_1,\vek{p}_2)$ and
$\dot{\vek{x}}_2=f(\vek{x}_1,\vek{x}_2,\vek{p}_1,\vek{p}_2)$ as given by the
Eqs.~\eqref{eq:x1punktUNDx2punktResultat}.
It is worth stressing that this method does not need the explicit knowledge of
the Lagrangian. We can write indeed:
\begin{subequations}
\label{eq:x1UNDx2punktpunktDefinition}
\begin{align}
\label{eq:x1punktpunktDefinition}
\ddot{\vek{x}}_1 = &\frac{d}{dt}\dot{\vek{x}}_1=
\frac{\pa \dot{\vek{x}}_1}{\pa \vek{x}_1}\dot{\vek{x}}_1
+\frac{\pa \dot{\vek{x}}_1}{\pa \vek{x}_2}\dot{\vek{x}}_2
+\frac{\pa \dot{\vek{x}}_1}{\pa \vek{p}_1}\dot{\vek{p}}_1
+\frac{\pa \dot{\vek{x}}_1}{\pa \vek{p}_2}\dot{\vek{p}}_2\,,
\\
\label{eq:x2punktpunktDefinition}
\ddot{\vek{x}}_2 =&\frac{d}{dt}\dot{\vek{x}}_2=
\frac{\pa \dot{\vek{x}}_2}{\pa \vek{x}_1}\dot{\vek{x}}_1
+\frac{\pa \dot{\vek{x}}_2}{\pa \vek{x}_2}\dot{\vek{x}}_2
+\frac{\pa \dot{\vek{x}}_2}{\pa \vek{p}_1}\dot{\vek{p}}_1
+\frac{\pa \dot{\vek{x}}_2}{\pa \vek{p}_2}\dot{\vek{p}}_2\,.
\end{align}
\end{subequations}

Inserting appropriate PN contributions to 
$\dot{\vek{x}}_a$ and $\dot{\vek{p}}_a$ given in
Eqs.~\eqref{eq:x1punktUNDx2punktResultat} and
\eqref{eq:p1punktUNDp2punktResultat}, we obtain
$1$PN , $2.5$PN and $3.5$PN corrections to the familiar Newtonian
expression for $\ddot{\vek{x}}_a$ and it reads --- apart from
the $2$PN and $3$PN terms --- \cite{CK2002}:
\begin{subequations}
\label{eq:x1UNDx2punktpunktinPResultat}
\begin{align}
\ddot{\vek{x}}_1 =\,
&-\frac{G\,m_2}{r^2}\,\vek{n}
+\frac{1}{c^2}
\Bigg\{
\frac{G}{r^2}
\bigg\{
\left[
-\frac{m_2}{m_1^2}\vek{p}_1^2
+\frac{4}{m_1}\left(\vek{p}_1\cdot\vek{p}_2\right)
+\frac{1}{m_2}
\left(
\frac{3}{2}\left(\vek{n}\cdot\vek{p}_2\right)^2-2\vek{p}_2^2
\right)
\right]\vek{n}
+\left[
4\frac{m_2}{m_1^2}\left(\vek{n}\cdot\vek{p}_1\right)
\right.
\nonumber\\[1mm]&
\left.
-3\frac{1}{m_1}\left(\vek{n}\cdot\vek{p}_2\right)
\right]\vek{p}_1
+\left[
-4\frac{1}{m_1}\left(\vek{n}\cdot\vek{p}_1\right)
+3\frac{1}{m_2}\left(\vek{n}\cdot\vek{p}_2\right)
\right]\vek{p}_2
\bigg\}
+\frac{G^2}{r^3}\bigg[5 m_1\,m_2 + 4 m_2^2\bigg]\vek{n}
\Bigg\}
\nonumber\\[1mm]&
+\frac{1}{c^5}
\Bigg\{
\frac{G^2}{r^3}
\bigg\{
\left[
\frac{m_2}{m_1^2}
\left(
-24\left(\vek{n}\cdot\vek{p}_1\right)^3
+\frac{96}{5}\left(\vek{n}\cdot\vek{p}_1\right)\vek{p}_1^2
\right)
\right.
+\frac{1}{m_1}
\left(
48\left(\vek{n}\cdot\vek{p}_1\right)^2\left(\vek{n}\cdot\vek{p}_2\right)
-\frac{72}{5}\left(\vek{n}\cdot\vek{p}_2\right)\vek{p}_1^2
\right.
\nonumber\\[1mm]&
-24\left(\vek{n}\cdot\vek{p}_1\right)\left(\vek{p}_1\cdot\vek{p}_2\right)
\bigg)
\left.
+\frac{1}{m_2}
\left(
\frac{72}{5}\left(\vek{n}\cdot\vek{p}_2\right)
\left(\vek{p}_1\cdot\vek{p}_2\right)
+\left(\vek{n}\cdot\vek{p}_1\right)
\left(
-24\left(\vek{n}\cdot\vek{p}_2\right)^2
+\frac{24}{5}\vek{p}_2^2
\right)
\right)
\right]\vek{n}
\nonumber\\[1mm]&
+\left[
\frac{m_2}{m_1^2}
\left(
\frac{64}{5}\left(\vek{n}\cdot\vek{p}_1\right)^2
-\frac{88}{15}\vek{p}_1^2
\right)
+\frac{1}{m_1}
\left(
-\frac{56}{5}\left(\vek{n}\cdot\vek{p}_1\right)\left(\vek{n}\cdot\vek{p}_2
\right)
+\frac{16}{3}\left(\vek{p}_1\cdot\vek{p}_2\right)
\right)
+\frac{1}{m_2}\left(
-\frac{8}{5}\left(\vek{n}\cdot\vek{p}_2\right)^2
\right.
\right.
\nonumber\\[1mm]&
\left.
\left.
+\frac{8}{15}\vek{p}_2^2
\right)
\right]\vek{p}_1
+\left[
\frac{1}{m_1}
\left(
-\frac{72}{5}\left(\vek{n}\cdot\vek{p}_1\right)^2
+\frac{32}{5}\vek{p}_1^2
\right)
+\frac{1}{m_2}
\left(
\frac{72}{5}\left(\vek{n}\cdot\vek{p}_1\right)\left(\vek{n}\cdot\vek{p}_2
\right)
-\frac{32}{5}\left(\vek{p}_1\cdot\vek{p}_2\right)
\right)
\right]\vek{p}_2
\bigg\}
\nonumber\\[1mm]&
+\frac{G^3}{r^4}
\bigg\{
\left[
\frac{16}{5} m_2^2\left(\vek{n}\cdot\vek{p}_1\right)
+m_1\,m_2\left(
\frac{8}{5}\left(\vek{n}\cdot\vek{p}_1\right)
-\frac{8}{5}\left(\vek{n}\cdot\vek{p}_2\right)
\right)
\right]\vek{n}
-\left[
\frac{8}{15}m_1\,m_2
+\frac{8}{15}m_2^2
\right]\vek{p}_1
\bigg\}
\nonumber\\[1mm]&
+\frac{1}{c^7}
\Bigg\{
\frac{G^2}{r^3}
\bigg\{
\left[
\frac{m_2}{m_1^4}
\left(
-46\left(\vek{n}\cdot\vek{p}_1\right)^5
+92\left(\vek{n}\cdot\vek{p}_1\right)^3\vek{p}_1^2
-\frac{1146}{35}\left(\vek{n}\cdot\vek{p}_1\right)\vek{p}_1^4
\right)
+\frac{1}{m_1^3}
\left(
214\left(\vek{n}\cdot\vek{p}_1\right)^4\left(\vek{n}\cdot\vek{p}_2\right)
\right.
\right.
\nonumber\\[1mm]&
\left.
-\frac{1756}{7}\left(\vek{n}\cdot\vek{p}_1\right)^2\left(\vek{n}\cdot
\vek{p}_2\right)\vek{p}_1^2
+\frac{934}{35}\left(\vek{n}\cdot\vek{p}_2\right)\vek{p}_1^4
-\frac{536}{7}\left(\vek{n}\cdot\vek{p}_1\right)^3\left(\vek{p}_1\cdot
\vek{p}_2\right)
+\frac{1772}{35}\left(\vek{n}\cdot\vek{p}_1\right)\vek{p}_1^2\left(\vek{p}_1
\cdot\vek{p}_2\right)
\right)
\nonumber\\[1mm]&
+\frac{1}{m_1^2\,m_2}
\left(
\frac{1452}{7}\left(\vek{n}\cdot\vek{p}_1\right)^2\left(\vek{n}\cdot
\vek{p}_2\right)\left(\vek{p}_1\cdot\vek{p}_2\right)
-44\left(\vek{n}\cdot\vek{p}_2\right)\vek{p}_1^2\left(\vek{p}_1\cdot\vek{p}_2
\right)
\right.
+\left(
-282\left(\vek{n}\cdot\vek{p}_2\right)^2
+\frac{46}{7}\vek{p}_2^2
\right)\left(\vek{n}\cdot\vek{p}_1\right)^3
\nonumber\\[1mm]&
\left.
+\left(
\frac{1322}{7}\left(\vek{n}\cdot\vek{p}_2\right)^2\vek{p}_1^2
-\frac{120}{7}\left(\vek{p}_1\cdot\vek{p}_2\right)^2
-10\vek{p}_1^2 \,\vek{p}_2^2 \right)\left(\vek{n}\cdot\vek{p}_1\right)
\right)
+\frac{1}{m_1\,m_2^2}
\bigg(
-42\left(\vek{n}\cdot\vek{p}_2\right)^3\vek{p}_1^2
\nonumber\\[1mm]&
+\left(\vek{n}\cdot\vek{p}_1\right)^2
\Big(
106\left(\vek{n}\cdot\vek{p}_2\right)^3
-2\left(\vek{n}\cdot\vek{p}_2\right)\vek{p}_2^2
\Big)
+\left(
\frac{816}{35}\left(\vek{p}_1\cdot\vek{p}_2\right)^2
+\frac{398}{35}\vek{p}_1^2\vek{p}_2^2
\right)\left(\vek{n}\cdot\vek{p}_2\right)
\nonumber\\[1mm]&
\left.
+\left(
-\frac{1032}{7}\left(\vek{n}\cdot\vek{p}_2\right)^2\left(\vek{p}_1\cdot
\vek{p}_2\right)
+\frac{232}{35}\left(\vek{p}_1\cdot\vek{p}_2\right)\vek{p}_2^2
\right)\left(\vek{n}\cdot\vek{p}_1\right)
\right)
+\frac{1}{m_2^3}
\left(
\frac{284}{7}\left(\vek{n}\cdot\vek{p}_2\right)^3\left(\vek{p}_1\cdot
\vek{p}_2\right)
\right.
\nonumber\\[1mm]&
-\frac{608}{35}\left(\vek{n}\cdot\vek{p}_2\right)\left(\vek{p}_1\cdot
\vek{p}_2\right)\vek{p}_2^2
\left.
\left.
+\left(
8\left(\vek{n}\cdot\vek{p}_2\right)^4
-\frac{116}{7}\left(\vek{n}\cdot\vek{p}_2\right)^2\vek{p}_2^2
+\frac{92}{35}\vek{p}_2^4
\right)\left(\vek{n}\cdot\vek{p}_1\right)
\right)
\right]\vek{n}
\nonumber\\[1mm]&
+\left[
\frac{m_2}{m_1^4}
\left(
\frac{334}{7}\left(\vek{n}\cdot\vek{p}_1\right)^4
-\frac{3028}{35}\left(\vek{n}\cdot\vek{p}_1\right)^2\vek{p}_1^2
+\frac{386}{21}\vek{p}_1^4
\right)
\right.
+\frac{1}{m_1^3}
\left(
-\frac{752}{7}\left(\vek{n}\cdot\vek{p}_1\right)^3\left(\vek{n}\cdot
\vek{p}_2\right)
\right.
\nonumber\\[1mm]&
+\frac{828}{7}\left(\vek{n}\cdot\vek{p}_1\right)\left(\vek{n}\cdot\vek{p}_2
\right)\vek{p}_1^2
\left.
+\frac{264}{5}\left(\vek{n}\cdot\vek{p}_1\right)^2\left(\vek{p}_1\cdot
\vek{p}_2\right)
-\frac{984}{35}\vek{p}_1^2\left(\vek{p}_1\cdot\vek{p}_2\right)
\right)
+\frac{1}{m_2^3}
\left(
-\frac{208}{7}\left(\vek{n}\cdot\vek{p}_2\right)^4
\right.
\nonumber\\[1mm]&
\left.
+\frac{1436}{35}\left(\vek{n}\cdot\vek{p}_2\right)^2\vek{p}_2^2
-\frac{116}{15}\vek{p}_2^4
\right)
+\frac{1}{m_1^2\,m_2}
\left(
-\frac{214}{7}\left(\vek{n}\cdot\vek{p}_2\right)^2\vek{p}_1^2
-\frac{1296}{35}\left(\vek{n}\cdot\vek{p}_1\right)\left(\vek{n}\cdot
\vek{p}_2\right)\left(\vek{p}_1\cdot\vek{p}_2\right)
\right.
\nonumber\\[1mm]&
\left.
+\frac{40}{7}\left(\vek{p}_1\cdot\vek{p}_2\right)^2
-\frac{10}{21}\vek{p}_1^2 \, \vek{p}_2^2
+\left(
6\left(\vek{n}\cdot\vek{p}_2\right)^2
+\frac{814}{35}\vek{p}_2^2
\right)\left(\vek{n}\cdot\vek{p}_1\right)^2
\right)
+\frac{1}{m_1\,m_2^2}
\left(
-\frac{60}{7}\left(\vek{n}\cdot\vek{p}_2\right)^2\left(\vek{p}_1\cdot
\vek{p}_2\right)
\right.
\nonumber\\[1mm]&
+\frac{428}{35}\left(\vek{p}_1\cdot\vek{p}_2\right)\vek{p}_2^2
\left.
\left.
+\left(
\frac{584}{7}\left(\vek{n}\cdot\vek{p}_2\right)^3
-\frac{2544}{35}\left(\vek{n}\cdot\vek{p}_2\right)\vek{p}_2^2
\right)\left(\vek{n}\cdot\vek{p}_1\right)
\right)
\right]\vek{p}_1
+\left[
\frac{1}{m_1^3}
\left(-82\left(\vek{n}\cdot\vek{p}_1\right)^4
\right.
\right.
\nonumber\\[1mm]&
\left.
+\frac{4416}{35}\left(\vek{n}\cdot\vek{p}_1\right)^2\vek{p}_1^2
-\frac{838}{35}\vek{p}_1^4
\right)
+\frac{1}{m_1^2\,m_2}
\left(
\frac{1684}{7}\left(\vek{n}\cdot\vek{p}_1\right)^3\left(\vek{n}\cdot\vek{p}_2
\right)
-\frac{7004}{35}\left(\vek{n}\cdot\vek{p}_1\right)\left(\vek{n}\cdot\vek{p}_2
\right)\vek{p}_1^2
\right.
\nonumber\\[1mm]&
\left.
-\frac{4852}{35}\left(\vek{n}\cdot\vek{p}_1\right)^2
\left(\vek{p}_1\cdot\vek{p}_2\right)
+\frac{396}{7}\vek{p}_1^2\left(\vek{p}_1\cdot\vek{p}_2\right)
\right)
+\frac{1}{m_1\,m_2^2}
\left(
\frac{2542}{35}\left(\vek{n}\cdot\vek{p}_2\right)^2\vek{p}_1^2
\right.
\nonumber\\[1mm]&
\left.
+\frac{1424}{7}\left(\vek{n}\cdot\vek{p}_1\right)\left(\vek{n}\cdot\vek{p}_2
\right)\left(\vek{p}_1\cdot\vek{p}_2\right)
-\frac{248}{7}\left(\vek{p}_1\cdot\vek{p}_2\right)^2
-\frac{622}{35}\vek{p}_1^2 \, \vek{p}_2^2
+\left(
-\frac{1394}{7}\left(\vek{n}\cdot\vek{p}_2\right)^2
+\frac{898}{35}\vek{p}_2^2
\right)\left(\vek{n}\cdot\vek{p}_1\right)^2
\right)
\nonumber\\[1mm]&
+\frac{1}{m_2^3}
\left(
-\frac{2512}{35}\left(\vek{n}\cdot\vek{p}_2\right)^2\left(\vek{p}_1\cdot
\vek{p}_2\right)
+\frac{144}{7}\left(\vek{p}_1\cdot\vek{p}_2\right)\vek{p}_2^2
\left.
+\left(
\frac{284}{7}\left(\vek{n}\cdot\vek{p}_2\right)^3
-\frac{608}{35}\left(\vek{n}\cdot\vek{p}_2\right)\vek{p}_2^2
\right)\left(\vek{n}\cdot\vek{p}_1\right)
\right)
\right]\vek{p}_2
\bigg\}
\nonumber\\[1mm]&
+\frac{G^3}{r^4}
\bigg\{
\left[
-\frac{5316}{7}\left(\vek{n}\cdot\vek{p}_1\right)^2\left(\vek{n}\cdot
\vek{p}_2\right)
+\frac{m_2^2}{m_1^2}
\left(
\frac{18084}{35}\left(\vek{n}\cdot\vek{p}_1\right)^3
-\frac{2472}{7}\left(\vek{n}\cdot\vek{p}_1\right)\vek{p}_1^2
\right)
\right.
+\left(
\frac{19984}{105}\vek{p}_1^2
\right.
\nonumber\\[1mm]&
\left.
-\frac{45592}{105}\left(\vek{p}_1\cdot\vek{p}_2\right)
\right)\left(\vek{n}\cdot\vek{p}_2\right)
+\frac{m_2}{m_1}
\left(
\frac{13476}{35}\left(\vek{n}\cdot\vek{p}_1\right)^3
-\frac{43208}{35}\left(\vek{n}\cdot\vek{p}_1\right)^2\left(\vek{n}\cdot
\vek{p}_2\right)
+\frac{31184}{105}\left(\vek{n}\cdot\vek{p}_2\right)\vek{p}_1^2
\right.
\nonumber\\[1mm]&
\left.
+\left(
-\frac{3916}{15}\vek{p}_1^2
+\frac{56752}{105}\left(\vek{p}_1\cdot\vek{p}_2\right)
\right)\left(\vek{n}\cdot\vek{p}_1\right)
\right)
+\left(
\frac{6284}{7}\left(\vek{n}\cdot\vek{p}_2\right)^2
+\frac{952}{3}\left(\vek{p}_1\cdot\vek{p}_2\right)
-\frac{19276}{105}\vek{p}_2^2
\right)\left(\vek{n}\cdot\vek{p}_1\right)
\nonumber\\[1mm]&
+\frac{m_1}{m_2}
\left(
-\frac{6464}{35}\left(\vek{n}\cdot\vek{p}_2\right)^3
+\left(
\frac{1848}{5}\left(\vek{n}\cdot\vek{p}_2\right)^2
-\frac{6716}{105}\vek{p}_2^2
\right)\left(\vek{n}\cdot\vek{p}_1\right)
\left.
+\left(
-184\left(\vek{p}_1\cdot\vek{p}_2\right)
+\frac{13868}{105}\vek{p}_2^2
\right)\left(\vek{n}\cdot\vek{p}_2\right)
\right)
\right]\vek{n}
\nonumber\\[1mm]&
+\left[
\frac{6724}{35}\left(\vek{n}\cdot\vek{p}_1\right)
\left(\vek{n}\cdot\vek{p}_2\right)
-\frac{1768}{15}\left(\vek{n}\cdot\vek{p}_2\right)^2
-\frac{2104}{35}\left(\vek{p}_1\cdot\vek{p}_2\right)
+\frac{4366}{105}\vek{p}_2^2
\right.
+\frac{m_2^2}{m_1^2}
\left(
-\frac{9876}{35}\left(\vek{n}\cdot\vek{p}_1\right)^2
+\frac{2874}{35}\vek{p}_1^2
\right)
\nonumber\\[1mm]&
+\frac{m_2}{m_1}
\left(
-\frac{20264}{105}\left(\vek{n}\cdot\vek{p}_1\right)^2
+\frac{6124}{15}\left(\vek{n}\cdot\vek{p}_1\right)\left(\vek{n}\cdot
\vek{p}_2\right)
+\frac{872}{15}\vek{p}_1^2
-\frac{13172}{105}\left(\vek{p}_1\cdot\vek{p}_2\right)
\right)
+\frac{m_1}{m_2}
\left(
-\frac{8}{15}\left(\vek{n}\cdot\vek{p}_2\right)^2
\right.
\nonumber\\[1mm]&
\left.
\left.
+\frac{596}{105}\vek{p}_2^2
\right)
\right]\vek{p}_1
+\left[
\frac{19604}{105}\left(\vek{n}\cdot\vek{p}_1\right)^2
-\frac{43328}{105}\left(\vek{n}\cdot\vek{p}_1\right)\left(\vek{n}\cdot
\vek{p}_2\right)
-\frac{348}{7}\vek{p}_1^2
+\frac{12356}{105}\left(\vek{p}_1\cdot\vek{p}_2\right)
+\frac{m_2}{m_1}
\left(
\frac{30052}{105}\left(\vek{n}\cdot\vek{p}_1\right)^2
\right.
\right.
\nonumber\\[1mm]&
\left.\left.
-\frac{2734}{35}\vek{p}_1^2
\right)
+\frac{m_1}{m_2}
\left(
-184\left(\vek{n}\cdot\vek{p}_1\right)\left(\vek{n}\cdot\vek{p}_2\right)
+\frac{12548}{105}\left(\vek{n}\cdot\vek{p}_2\right)^2
+\frac{704}{15}\left(\vek{p}_1\cdot\vek{p}_2\right)
-\frac{3874}{105}\vek{p}_2^2
\right)
\right]\vek{p}_2
\bigg\}
\nonumber\\[1mm]&
+\frac{G^4}{r^5}
\bigg\{
\left[
-\frac{296}{15} m_2^3\left(\vek{n}\cdot\vek{p}_1\right)
+m_1^2\,m_2
\left(
\frac{8}{35}\left(\vek{n}\cdot\vek{p}_1\right)
\right.
\left.
+\frac{344}{15}\left(\vek{n}\cdot\vek{p}_2\right)
\right)
+m_1\,m_2^2
\left(
-\frac{872}{35}\left(\vek{n}\cdot\vek{p}_1\right)
+\frac{352}{15}\left(\vek{n}\cdot\vek{p}_2\right)
\right)
\right]\vek{n}
\nonumber\\[1mm]&
+\left[
-\frac{8}{105} m_1^2\,m_2
-\frac{226}{105}m_1\,m_2^2
-\frac{218}{105}m_2^3
\right]\vek{p}_1
+\left[
\frac{58}{21}m_1^2\,m_2
+\frac{62}{35}m_1\,m_2^2
\right]\vek{p}_2
\bigg\}
\Bigg\}
\,,
\label{eq:x1punktpunktinPResultat}
\\[1mm]
\label{eq:x2punktpunktinPResultat}
\ddot{\vek{x}}_2 = &\; (1 \rightleftharpoons 2)
\,.
\end{align}
\end{subequations}

We now go on to present $\ddot{\vek{x}}_a$ in terms of
($\vek{x}_a$,$\vek{v}_a$). As Eqs.~\eqref{eq:x1UNDx2punktpunktinPResultat}
involve $\vek{p}_a$ at $1$PN, $2.5$PN and $3.5$PN orders, we use only
Newtonian, $1$PN, $2.5$PN contributions to $\vek{p}_a$, given in 
Eqs.~\eqref{eq:p1UNDp2ausInvertierung} [rather than
Eqs.~\eqref{eq:changeP1andP2toV} truncated at the $1$PN approximation].
After some heavy algebra, we get
$\ddot{\vek{x}}_1=f(\vek{x}_1,\vek{x}_2,\vek{v}_1,\vek{v}_2)$
and
$\ddot{\vek{x}}_2=f(\vek{x}_1,\vek{x}_2,\vek{v}_1,\vek{v}_2)$
\cite{CK2002}:
\begin{subequations}
\label{eq:x1UNDx2pupuResultat}
\begin{align}
\ddot{\vek{x}}_1 =&\; -\frac{G m_2}{r^2} \vek{n}
+\frac{1}{c^2}
\Bigg\{
\frac{G m_2}{r^2}
\bigg\{
\left[
\frac{3}{2}\left(\vek{n}\cdot\vek{v}_2\right)^2
-\vek{v}_1^2
+4\left(\vek{v}_1\cdot\vek{v}_2\right)
-2\vek{v}_2^2
\right]
\vek{n}
+\bigg[
4\left(\vek{n}\cdot\vek{v}_1\right)
-3\left(\vek{n}\cdot\vek{v}_2\right)
\bigg]
\left(\vek{v}_1-\vek{v}_2\right)
\bigg\}
\nonumber\\[1mm]&
+\frac{G^2 m_2}{r^3}\bigg[5 m_1 + 4 m_2\bigg]\vek{n}
\Bigg\}
+\frac{1}{c^5}
\Bigg\{
\frac{G^2 m_1 m_2}{r^3}
\bigg\{
\left[
-24\left(\vek{n}\cdot\vek{v}_1\right)^3
+\left(
48\left(\vek{n}\cdot\vek{v}_1\right)^2
-\frac{72}{5}\vek{v}_1^2
+\frac{72}{5}\left(\vek{v}_1\cdot\vek{v}_2\right)
\right)\left(\vek{n}\cdot\vek{v}_2\right)
\right.
\nonumber\\[1mm]&
\left.
+\left(
-24\left(\vek{n}\cdot\vek{v}_2\right)^2
+\frac{96}{5}\vek{v}_1^2
-24\left(\vek{v}_1\cdot\vek{v}_2\right)
+\frac{24}{5}\vek{v}_2^2\right)\left(\vek{n}\cdot\vek{v}_1\right)
\right]
\vek{n}
+\left[
\frac{64}{5}\left(\vek{n}\cdot\vek{v}_1\right)^2
-\frac{56}{5}\left(\vek{n}\cdot\vek{v}_1\right)\left(\vek{n}\cdot\vek{v}_2
\right)
\right.
\nonumber\\[1mm]&
\left.
-\frac{8}{5}\left(\vek{n}\cdot\vek{v}_2\right)^2
-\frac{88}{15}\vek{v}_1^2
+\frac{16}{3}\left(\vek{v}_1\cdot\vek{v}_2\right)
+\frac{8}{15}\vek{v}_2^2
\right]\vek{v}_1
+\left[
-\frac{72}{5}\left(\vek{n}\cdot\vek{v}_1\right)^2
+\frac{72}{5}\left(\vek{n}\cdot\vek{v}_1\right)\left(\vek{n}\cdot\vek{v}_2
\right)
+\frac{32}{5}\vek{v}_1^2
\right.
\nonumber\\[1mm]&
\left.
-\frac{32}{5}\left(\vek{v}_1\cdot\vek{v}_2\right)
\right]
\vek{v}_2
\bigg\}
+\frac{G^3 m_1 m_2}{r^4}
\bigg\{
\left[
\frac{8}{5} m_1 \left(\vek{n}\cdot\vek{v}_1\right)
+m_2\left(\frac{16}{5}\left(\vek{n}\cdot\vek{v}_1\right)
-\frac{8}{5}\left(\vek{n}\cdot\vek{v}_2\right)\right)
\right]\vek{n}
-\left[
\frac{8}{15}m_1
+\frac{8}{15}m_2
\right]
\vek{v}_1
\bigg\}
\Bigg\}
\nonumber\\[1mm]&
+\frac{1}{c^7}
\Bigg\{
\frac{G^2 m_1 m_2}{r^3}
\bigg\{
\bigg[
-46\left(\vek{n}\cdot\vek{v}_1\right)^5
+214\left(\vek{n}\cdot\vek{v}_1\right)^4\left(\vek{n}\cdot\vek{v}_2\right)
+106\left(\vek{n}\cdot\vek{v}_1\right)^2\left(\vek{n}\cdot\vek{v}_2\right)^3
+\bigg(
-42\vek{v}_1^2
\nonumber\\[1mm]&
\left.
+\frac{284}{7}\left(\vek{v}_1\cdot\vek{v}_2\right)
\right)\left(\vek{n}\cdot\vek{v}_2\right)^3
+\left(
\frac{1238}{7}\vek{v}_1^2
-\frac{1032}{7}\left(\vek{v}_1\cdot\vek{v}_2\right)
-\frac{284}{7}\vek{v}_2^2
\right)\left(\vek{n}\cdot\vek{v}_1\right)
\left(\vek{n}\cdot\vek{v}_2\right)^2
+\bigg(
-282\left(\vek{n}\cdot\vek{v}_2\right)^2
\nonumber\\[1mm]&
\left.
+56\vek{v}_1^2
-\frac{536}{7}\left(\vek{v}_1\cdot\vek{v}_2\right)
+\frac{46}{7}\vek{v}_2^2
\right)\left(\vek{n}\cdot\vek{v}_1\right)^3
+\left(
-\frac{1420}{7}\vek{v}_1^2
+\frac{1452}{7}\left(\vek{v}_1\cdot\vek{v}_2\right)
+22\vek{v}_2^2
\right)\left(\vek{n}\cdot\vek{v}_1\right)^2\left(\vek{n}\cdot\vek{v}_2\right)
\nonumber\\[1mm]&
+\left(
\frac{86}{7}\vek{v}_1^4
-\frac{184}{5}\vek{v}_1^2\left(\vek{v}_1\cdot\vek{v}_2\right)
+\frac{816}{35}\left(\vek{v}_1\cdot\vek{v}_2\right)^2
+\frac{146}{35}\vek{v}_1^2\vek{v}_2^2
-\frac{104}{35}\left(\vek{v}_1\cdot\vek{v}_2\right)\vek{v}_2^2
\right)\left(\vek{n}\cdot\vek{v}_2\right)
+\bigg(
8\left(\vek{n}\cdot\vek{v}_2\right)^4
\nonumber\\[1mm]&
\left.
\left.
-\frac{138}{35}\vek{v}_1^4
+\frac{932}{35}\vek{v}_1^2\left(\vek{v}_1\cdot\vek{v}_2\right)
-\frac{120}{7}\left(\vek{v}_1\cdot\vek{v}_2\right)^2
-\frac{38}{5}\vek{v}_1^2\vek{v}_2^2
-\frac{188}{35}\left(\vek{v}_1\cdot\vek{v}_2\right)\vek{v}_2^2
+\frac{52}{7}\vek{v}_2^4
\right)\left(\vek{n}\cdot\vek{v}_1\right)
\right]
\vek{n}
\nonumber\\[1mm]&
+\left[
\frac{334}{7}\left(\vek{n}\cdot\vek{v}_1\right)^4
-\frac{752}{7}\left(\vek{n}\cdot\vek{v}_1\right)^3\left(\vek{n}\cdot\vek{v}_2
\right)
\right.
+\frac{584}{7}\left(\vek{n}\cdot\vek{v}_1\right)
{\left(\vek{n}\cdot\vek{v}_2\right)}^3
-\frac{208}{7}\left(\vek{n}\cdot\vek{v}_2\right)^4
+\frac{1006}{105}\vek{v}_1^4
-\frac{2392}{105}\vek{v}_1^2\left(\vek{v}_1\cdot\vek{v}_2\right)
\nonumber\\[1mm]&
+\frac{40}{7}\left(\vek{v}_1\cdot\vek{v}_2\right)^2
+\left(
\frac{3748}{35}\vek{v}_1^2
-\frac{1296}{35}\left(\vek{v}_1\cdot\vek{v}_2\right)
-\frac{548}{7}\vek{v}_2^2
\right)\left(\vek{n}\cdot\vek{v}_1\right)\left(\vek{n}\cdot\vek{v}_2\right)
-\frac{22}{105}\vek{v}_1^2\vek{v}_2^2
+\frac{1564}{105}\left(\vek{v}_1\cdot\vek{v}_2\right)\vek{v}_2^2
-\frac{36}{5}\vek{v}_2^4
\nonumber\\[1mm]&
+\left(
6\left(\vek{n}\cdot\vek{v}_2\right)^2
-\frac{2356}{35}\vek{v}_1^2
+\frac{264}{5}\left(\vek{v}_1\cdot\vek{v}_2\right)
+\frac{814}{35}\vek{v}_2^2
\right)\left(\vek{n}\cdot\vek{v}_1\right)^2
\left.
+\left(
-\frac{1098}{35}\vek{v}_1^2
-\frac{60}{7}\left(\vek{v}_1\cdot\vek{v}_2\right)
+\frac{276}{7}\vek{v}_2^2
\right)\left(\vek{n}\cdot\vek{v}_2\right)^2
\right]
\vek{v}_1
\nonumber\\[1mm]&
+\left[
-82\left(\vek{n}\cdot\vek{v}_1\right)^4
+\frac{1684}{7}\left(\vek{n}\cdot\vek{v}_1\right)^3\left(\vek{n}\cdot
\vek{v}_2\right)
+\frac{284}{7}\left(\vek{n}\cdot\vek{v}_1\right)\left(\vek{n}\cdot
\vek{v}_2\right)^3
\right.
+\left(
\frac{2542}{35}\vek{v}_1^2
-\frac{2512}{35}\left(\vek{v}_1\cdot\vek{v}_2\right)
\right)\left(\vek{n}\cdot\vek{v}_2\right)^2
\nonumber\\[1mm]&
-\frac{614}{35}\vek{v}_1^4
+\frac{1868}{35}\vek{v}_1^2\left(\vek{v}_1\cdot\vek{v}_2\right)
-\frac{248}{7}\left(\vek{v}_1\cdot\vek{v}_2\right)^2
-\frac{102}{7}\vek{v}_1^2\vek{v}_2^2
+\frac{496}{35}\left(\vek{v}_1\cdot\vek{v}_2\right)\vek{v}_2^2
+\left(
-\frac{6752}{35}\vek{v}_1^2
+\frac{1424}{7}\left(\vek{v}_1\cdot\vek{v}_2\right)
\right.
\nonumber\\[1mm]&
\left.
\left.
-\frac{104}{35}\vek{v}_2^2
\right)\left(\vek{n}\cdot\vek{v}_1\right)\left(\vek{n}\cdot\vek{v}_2\right)
+\left(
-\frac{1394}{7}\left(\vek{n}\cdot\vek{v}_2\right)^2
+\frac{3912}{35}\vek{v}_1^2
-\frac{4852}{35}\left(\vek{v}_1\cdot\vek{v}_2\right)
+\frac{646}{35}\vek{v}_2^2
\right)\left(\vek{n}\cdot\vek{v}_1\right)^2
\right]
\vek{v}_2
\bigg\}
\nonumber\\[1mm]&
+\frac{G^3 m_1 m_2}{r^4}
\bigg\{
\left[
m_1\left(
\frac{7428}{35}\left(\vek{n}\cdot\vek{v}_1\right)^3
-\frac{15492}{35}\left(\vek{n}\cdot\vek{v}_1\right)^2\left(\vek{n}\cdot
\vek{v}_2\right)
\right.
\right.
+\left(
\frac{10156}{105}\vek{v}_1^2
-\frac{488}{5}\left(\vek{v}_1\cdot\vek{v}_2\right)
\right)\left(\vek{n}\cdot\vek{v}_2\right)
\nonumber\\[1mm]&
\left.
+\left(
\frac{1128}{5}\left(\vek{n}\cdot\vek{v}_2\right)^2
-\frac{1828}{15}\vek{v}_1^2
+\frac{472}{3}\left(\vek{v}_1\cdot\vek{v}_2\right)
-\frac{3692}{105}\vek{v}_2^2
\right)\left(\vek{n}\cdot\vek{v}_1\right)
\right)
+m_2\left(
\frac{10188}{35}\left(\vek{n}\cdot\vek{v}_1\right)^3
\right.
\nonumber\\[1mm]&
-\frac{23776}{35}\left(\vek{n}\cdot\vek{v}_1\right)^2\left(\vek{n}\cdot
\vek{v}_2\right)
-\frac{3272}{35}\left(\vek{n}\cdot\vek{v}_2\right)^3
+\left(
\frac{16692}{35}\left(\vek{n}\cdot\vek{v}_2\right)^2
-\frac{18656}{105}\vek{v}_1^2
+\frac{9472}{35}\left(\vek{v}_1\cdot\vek{v}_2\right)
-\frac{10064}{105}\vek{v}_2^2
\right)\left(\vek{n}\cdot\vek{v}_1\right)
\nonumber\\[1mm]&
\left.
\left.
+\left(
\frac{5308}{35}\vek{v}_1^2
-\frac{22576}{105}\left(\vek{v}_1\cdot\vek{v}_2\right)
+\frac{6896}{105}\vek{v}_2^2
\right)\left(\vek{n}\cdot\vek{v}_2\right)
\right)
\right]
\vek{n}
+\left[
m_1\left(
-\frac{3516}{35}\left(\vek{n}\cdot\vek{v}_1\right)^2
+\frac{2528}{21}\left(\vek{n}\cdot\vek{v}_1\right)
\left(\vek{n}\cdot\vek{v}_2
\right)
\right.
\right.
\nonumber\\[1mm]&
\left.
-\frac{152}{15}\left(\vek{n}\cdot\vek{v}_2\right)^2
+\frac{84}{5}\vek{v}_1^2
-\frac{2672}{105}\left(\vek{v}_1\cdot\vek{v}_2\right)
+\frac{932}{105}\vek{v}_2^2
\right)
+m_2\left(
-\frac{17308}{105}\left(\vek{n}\cdot\vek{v}_1\right)^2
+\frac{1204}{5}\left(\vek{n}\cdot\vek{v}_1\right)\left(\vek{n}\cdot\vek{v}_2
\right)
\right.
\nonumber \\[1mm] &
\left.
\left.
-\frac{1184}{15}\left(\vek{n}\cdot\vek{v}_2\right)^2
+\frac{4394}{105}\vek{v}_1^2
-\frac{7852}{105}\left(\vek{v}_1\cdot\vek{v}_2\right)
+\frac{1194}{35}\vek{v}_2^2
\right)
\right]
\vek{v}_1
+\left[
m_1\left(
\frac{9356}{105}\left(\vek{n}\cdot\vek{v}_1\right)^2
\right.
\right.
\nonumber\\[1mm]&
\left.
-\frac{488}{5}\left(\vek{n}\cdot\vek{v}_1\right)\left(\vek{n}\cdot\vek{v}_2
\right)
-\frac{284}{35}\vek{v}_1^2
+\frac{128}{15}\left(\vek{v}_1\cdot\vek{v}_2\right)
\right)
+m_2\left(
\frac{16052}{105}\left(\vek{n}\cdot\vek{v}_1\right)^2
-\frac{1524}{7}\left(\vek{n}\cdot\vek{v}_1\right)\left(\vek{n}\cdot\vek{v}_2
\right)
\right.
\nonumber\\[1mm]&
\left.
\left.
+\frac{7256}{105}\left(\vek{n}\cdot\vek{v}_2\right)^2
-\frac{3358}{105}\vek{v}_1^2
+\frac{6028}{105}\left(\vek{v}_1\cdot\vek{v}_2\right)
-\frac{2726}{105}\vek{v}_2^2
\right)
\right]
\vek{v}_2
\bigg\}
+\frac{G^4 m_1 m_2}{r^5}
\bigg\{
\left[
\frac{8}{35} m_1^2 \left(\vek{n}\cdot\vek{v}_1\right)
\right.
\nonumber\\[1mm]&
\left.
+m_2^2
\left(
-\frac{152}{15}\left(\vek{n}\cdot\vek{v}_1\right)
+\frac{164}{15}\left(\vek{n}\cdot\vek{v}_2\right)
\right)
+m_1 m_2
\left(
-\frac{96}{7}\left(\vek{n}\cdot\vek{v}_1\right)
+12\left(\vek{n}\cdot\vek{v}_2\right)
\right)
\right]
\vek{n}
\nonumber\\[1mm]&
+\left[
-\frac{8}{105} m_1^2
-\frac{394}{105} m_1 m_2
-\frac{386}{105} m_2^2
\right]
\vek{v}_1
+\left[
\frac{162}{35}m_1 m_2
+\frac{382}{105}m_2^2
\right]
\vek{v}_2
\bigg\}
\Bigg\}
\,,
\label{eq:x1pupuResultat}
\\[1mm]
\ddot{\vek{x}}_2 = &\;(1 \rightleftharpoons 2)
\,.
\label{eq:x2pupuResultat}
\end{align}
\end{subequations}
\end{widetext}
In the next subsection we show the second way to compute the equations
above.

\subsection{Second method to calculate the Euler-Lagrangian
equations of motions}

The second route is less straightforward, but leads to the confirmation of the
preceding result. We start from the Hamiltonian
equations~(\ref{eq:HamBewegungsGlDefinition}) for $\dot{\vek{p}}_a$:
\begin{align}
\label{eq:PaPunktDefinition}
\dot {\vek p}_a =-\frac{\partial H}{\partial \vek{x}_a}\,,
\end{align}
where we replace $\dot {\vek p}_a$ by the derivative of the right-hand side of
Eq.~\eqref{eq:p1UNDp2ausInvertierung}. The formal time derivatives of
$\dot{\vek{x}}_1$ and $\dot{\vek{x}}_2$, denoted by $\vek{a}_1$ and
$\vek{a}_2$ respectively, are kept unevaluated. The right-hand side of
Eq.~\eqref{eq:PaPunktDefinition} involve the reaction contributions
Eqs.~\eqref{eq:p1punktUNDp2punktResultat}, as well as
the Newtonian and post-Newtonian parts \eqref{eq:p1strichUNDp2strichNundPN}.
We systematically eliminate the particle
momenta in favor of the particle coordinate velocities by means of
Eqs.~\eqref{eq:p1UNDp2ausInvertierung} (without $1/c^7$).
To calculate $\ddot{\vek{x}}_1$, we consider
the coefficient of $\vek{a}_1$ in $\dot{\vek{p}}_1$, and invert it up to the
order $1/c^7$. The equation of motion \eqref{eq:PaPunktDefinition} in its
present form is multiplied by the latter result. Next, we gather all terms but
$\vek{a}_1$ in the right-hand side. Simultaneously, the acceleration
$\ddot{\vek{x}}_2$ is isolated in the same way. We arrive at the coupled
system:
\begin{subequations}
\begin{align}
\vek{a}_1 = &f(\vek{x}_1,\vek{x}_2,\dot{\vek{x}}_1
=\vek{v}_1,\dot{\vek{x}}_2
=\vek{v}_2,\ddot{\vek{x}}_1
=\vek{a}_1,\ddot{\vek{x}}_2=\vek{a}_2)
\,,\\
\vek{a}_2 = &f(\vek{x}_1,\vek{x}_2,\dot{\vek{x}}_1
=\vek{v}_1,\dot{\vek{x}}_2
=\vek{v}_2,\ddot{\vek{x}}_1
=\vek{a}_1,\ddot{\vek{x}}_2=\vek{a}_2) \,.
\end{align}
\end{subequations}
The various contributions are sorted in powers of $1/c$, and labeled with two
different indices. The second index refers to the $1/c$ exponent, whereas the
first index identifies the body; $N$ stands for the Newtonian part.
\begin{subequations}
\label{eq:a1UNDa2Wechselseitig}
\begin{align}
\label{eq:a1Wechselseitig}
\vek{a}_1=&\text{N}_{10}
+\frac{1}{c^2}{\rm f_{12}}(\vek{a}_1,\vek{a}_2)
+\frac{1}{c^5}{\rm f_{15}}(\vek{a}_1,\vek{a}_2)
+\frac{1}{c^7}{\rm f_{17}}(\vek{a}_1,\vek{a}_2)
\,,\\
\label{eq:a2Wechselseitig}
\vek{a}_2=&\text{N}_{20}
+\frac{1}{c^2}{\rm f_{22}}(\vek{a}_1,\vek{a}_2)
+\frac{1}{c^5}{\rm f_{25}}(\vek{a}_1,\vek{a}_2)
+\frac{1}{c^7}{\rm f_{27}}(\vek{a}_1,\vek{a}_2)
\,.
\end{align}
\end{subequations}
The decoupling is achieved by iterative substitution of
Eq.~\eqref{eq:a1Wechselseitig} into Eq.~\eqref{eq:a2Wechselseitig} and vice
versa. We get
\begin{subequations}
\label{eq:a1UNDa2ohneKopplungDef}
\begin{align}
\label{eq:a1ohneKopplungDef}
\vek{a}_1=&\text{N}_{10}
+\frac{1}{c^2}{\rm g_{12}}(\vek{a}_1)
+\frac{1}{c^5}{\rm g_{15}}(\vek{a}_1)
+\frac{1}{c^7}{\rm g_{17}}(\vek{a}_1)
\,,\\
\label{eq:a2ohneKopplungDef}
\vek{a}_2=&\text{N}_{20}
+\frac{1}{c^2}{\rm g_{22}}(\vek{a}_2)
+\frac{1}{c^5}{\rm g_{25}}(\vek{a}_2)
+\frac{1}{c^7}{\rm g_{27}}(\vek{a}_2)
\,.
\end{align}
\end{subequations}
Next, we compute $\vek{a}_1$ and $\vek{a}_2$ iteratively, and insert the
result into the right-hand sides of
Eqs.~\eqref{eq:a1UNDa2ohneKopplungDef}.
The acceleration has finally the following structure
\begin{subequations}
\begin{align}
\label{eq:a_1_ergebnis}
\vek{a}_1=&f(\vek{x}_1,\vek{x}_2,\dot{\vek{x}}_1=\vek{v}_1,\dot{\vek{x}}_2=
\vek{v}_2)
\,,
\\
\label{eq:a_2_ergebnis}
\vek{a}_2=&f(\vek{x}_1,\vek{x}_2,\dot{\vek{x}}_1=\vek{v}_1,\dot{\vek{x}}_2=
\vek{v}_2)
\,.
\end{align}
\end{subequations}
We recover Eqs.~\eqref{eq:x1UNDx2pupuResultat}.

In order to confirm the correctness of expressions above
for $\ddot{\vek{x}}_1$ and $\ddot{\vek{x}}_2$,
we shall check it yields the correct energy loss
\eqref{eq:zeitlgemttEloss420Version2}. For this goal, we begin by deriving the
reaction forces with the help of Eqs.~\eqref{eq:x1UNDx2pupuResultat}
\begin{subequations}
\begin{align}
\vek{F}_1^{\rm reac} &= m_1\,\ddot{\vek{x}}_1
\,,
\\
\vek{F}_2^{\rm reac} &= m_2\,\ddot{\vek{x}}_2
\,.
\end{align}
\end{subequations}
After multiplying these relations by $\dot{\vek{x}}_1=\vek{v}_1$ or 
$\dot{\vek{x}}_2=\vek{v}_2$ respectively, we sum to obtain
\be
\label{eq:Definition_von_Klein_h} 
m_1 \left( \ddot{\vek{x}}_1 \cdot \vek{v}_1 \right) +m_2 \left(
\ddot{\vek{x}}_2 \cdot \vek{v}_2 \right) =h(\vek{x}_1, \vek{x}_2,
\vek{v}_1, \vek{v}_2) 
\ee 
This is not yet the instantaneous energy loss due particularly to the presence
of the Newtonian and post-Newtonian terms coming from
Eqs.~\eqref{eq:x1UNDx2pupuResultat}. We then must
search for the function $g=g(\vek{x}_1, \vek{x}_2, \vek{v}_1, \vek{v}_2)$ to
be added to $h$ as defined in Eq.~\eqref{eq:Definition_von_Klein_h} in order
to isolate the radiative $2.5$PN and $3.5$PN parts, which does not only amount
to removing the Newtonian and post-Newtonian contributions. We shall then be
able to find the average power by substituting to $\vek{v}^2$ its
value~\eqref{eq:VVausEnergieformel335}.

The correct energy loss is given by the time derivative of
\mbox{$\mathcal{E}=H(\vek{x}_1, \vek{x}_2, \vek{p}_1 (\vek{x}_a,
\vek{v}_a), \vek{p}_2(\vek{x}_a, \vek{v}_a))$}
truncated at the $1$PN level. The particle momenta of
the $1$PN Hamiltonian~\eqref{eq:HAMILTONfkt-1PN-ADM} are expressed by means of
the particle coordinate velocities~\eqref{eq:p1UNDp2ausInvertierung} up
to the order $1/c^5$ (only). When performing the differentiation
of the resulting energy with respect to time,
$\ddot{\vek{x}}_1=\vek{a}_1$ and $\ddot{\vek{x}}_2=\vek{a}_2$ --- coming from
the derivatives of $\dot{\vek{x}}_1=\vek{v}_1$ and
$\dot{\vek{x}}_2=\vek{v}_2$ --- are replaced by
expressions~\eqref{eq:x1UNDx2pupuResultat} except in the Newtonian term
\mbox{$m_1 \left( \vek{a}_1 \cdot \vek{v}_1 \right) +m_2 \left( \vek{a}_2
\cdot \vek{v}_2 \right)$} where they are left unevaluated. The function
$g=g(\vek{x}_1, \vek{x}_2, \vek{v}_1, \vek{v}_2)$ is just the difference
\mbox{$d\mathcal{E}/dt - m_1 \left( \vek{a}_1 \cdot \vek{v}_1 \right)
- m_2 \left(\vek{a}_2 \cdot \vek{v}_2 \right)$}.

By construction, $h+g$ contains no Newtonian or post-Newtonian terms. It
actually represents the instantaneous luminosity
${\cal L}_{\le{\rm 3.5PN}}^{\text{(3)}}= - d\mathcal{E}/dt$,
we intended to check, given by the negative instantaneous energy
loss.\footnote{The full result for $\langle {\cal L}_{\le{\rm
3.5PN}}^{\text{(3)}} \rangle$ is presented in Ref.~\cite{CK2002}.}
The result of the averaging procedure in the case of quasi-elliptic
orbital motion is in agreement with our preceding
results and with Eqs.~(4.20) and (4.21) of Ref.~\cite{BS89}.

\section{Comparism with Iyer-Will formalism}

In this section, we shall show that the dissipative parts of
Eqs.~\eqref{eq:x1UNDx2pupuResultat}
are compatible with the generic reaction force of Iyer and Will
\cite{IW95} depending on $(2+6)$ gauge parameters,
as the ones given in the Eqs.~(4.1a)--(4.1d) in Ref.~\cite{IW95}.
The first two parameters at the $2.5$PN order and the six coefficients
at the $3.5$PN order correspond to the gauge freedom.

As a matter of fact, it is sufficient to prove that for certain values
of the unknowns, the latter reaction force coincides with ours.
This requires to compute the relative acceleration
\be
\label{eq:RelBeschlDef}
\vek{a}=\vek{a}_{12}:=\vek{a}_1-\vek{a}_2=\ddot{\vek{x}}_1-\ddot{\vek{x}}_2
\ee
from Eq.~(\ref{eq:x1UNDx2punktpunktinPResultat}) for 
$\ddot{\vek{x}}_1=f(\vek{x}_1,\vek{x}_2,\vek{p}_1,\vek{p}_2)$ 
and $\ddot{\vek{x}}_2=f(\vek{x}_1,\vek{x}_2,\vek{p}_1,\vek{p}_2)$
respectively and to go to the center-of-mass frame
\be
\vek{p}_1+\vek{p}_2=0
\,.
\ee
As the accelerations we start with are function of $\vek{p}_1$ and $\vek{p}_2$
rather than $\vek{v}_1$ and $\vek{v_2}$, we must simply set $\vek{p}_1$
to $\vek{p}$ and $\vek{p}_2$ to $-\vek{p}$
in $\vek{a}=\vek{a}(\vek{x}_1, \vek{x}_2, \vek{p}_1, \vek{p}_2)$
[see Eq.~\eqref{eq:RelBeschlDef}].

The next step consists in establishing the link between the particle momenta
$\vek{p}$ and the relative velocity
$\vek{v}=\vek{v}_{12}:=\vek{v}_1-\vek{v}_2$ in order to eliminate the particle
momenta from $\vek{a}=\vek{a}(\vek{x}_1, \vek{x}_2, \vek{p})$. We thus restore
the Newtonian and post-Newtonian part in
Eqs.~\eqref{eq:x1punktUNDx2punktResultat} and calculate the relative velocity 
\be
\vek{v}=\vek{v}_{12}:=\vek{v}_1-\vek{v}_2=\dot{\vek{x}}_1-\dot{\vek{x}}_2
\,.
\ee 
We make the substitutions $\vek{p}_1 \rightarrow \vek{p}$ and $\vek{p}_2
\rightarrow -\vek{p}$ in $\vek{v}=\vek{v}(\vek{x}_1, \vek{x}_2, \vek{p}_1,
\vek{p}_2)$. The relation $\vek{v}=\vek{v}(\vek{x}_1, \vek{x}_2, \vek{p})$ can
be inverted with the aid of the iterative method described above to get the
momentum as a function of the relative velocity $\vek{p}=\vek{p}(\vek{x}_1,
\vek{x}_2, \vek{v})$. Inserting $\vek{p}=\vek{p}(\vek{x}_1, \vek{x}_2,
\vek{v})$ in $\vek{a}=\vek{a}(\vek{x}_1, \vek{x}_2, \vek{p})$ leads to
$\vek{a}=\vek{a}(\vek{x}_1, \vek{x}_2, \vek{v})$. After introducing the total
mass $M$ as well as the dimensionless mass parameter $\nu$ through
Eqs.~\eqref{eq:RelMasses} and $\mu = \nu M$, the relative acceleration
reads \cite{CK2002}:
\begin{widetext}
\begin{align}
\vek{a}=& \mbox{}
-\frac{G M }{r^2}\vek{n}
+\frac{1}{c^2}
\Bigg\{
\frac{G M}{r^2}
\bigg\{
\left[
-\left(1+3 \nu\right)\vek{v}^2
+\frac{3}{2} \dot{r}^2\,\nu
\right]\vek{n}
+\left(
4 \dot{r}-2 \dot{r} \nu
\right)\vek{v}
\bigg\}
+\frac{G^2 M^2}{r^3} \left( 4+2 \nu \right)\vek{n}
\Bigg\}
\nonumber\\ 
&+\frac{1}{c^5}
\Bigg\{
\frac{G^2 M^2}{r^3}
\left[
\left(
-24 \dot{r}^3 \nu
+\frac{96}{5} \dot{r} \vek{v}^2 \nu
\right)\vek{n}
+\left(
\frac{64}{5} \dot{r}^2 \nu
-\frac{88}{15} \vek{v}^2 \nu
\right)\vek{v}
\right]
+\frac{G^3 M^3}{r^4}
\left(
\frac{16}{5} \dot{r} \nu \vek{n}
-\frac{8}{15} \nu \vek{v}
\right)
\Bigg\}
\nonumber\\ 
&+\frac{1}{c^7}
\Bigg\{
\frac{G^2 M^2}{r^3}
\bigg\{
\left[
-46 \dot{r}^5 \nu
+24 \dot{r}^5 \nu^2
+\vek{v}^4
\left(
-\frac{138}{35} \dot{r} \nu
-\frac{516}{35} \dot{r} \nu^2
\right)
+\vek{v}^2
\left(56 \dot{r}^3 \nu
-\frac{4}{7} \dot{r}^3 \nu^2
\right)
\right]\vek{n}
\nonumber\\
&+\left[
\frac{334}{7} \dot{r}^4 \nu
-\frac{268}{7} \dot{r}^4 \nu^2
+\vek{v}^4
\left(
\frac{1006}{105} \nu
-\frac{64}{105} \nu^2
\right)
+\vek{v}^2
\left(
-\frac{2356}{35} \dot{r}^2 \nu
+\frac{148}{5} \dot{r}^2 \nu^2
\right)
\right]\vek{v}
\bigg\}
\nonumber\\ 
&+\frac{G^3 M^3}{r^4}
\bigg\{
\left[
\frac{10188}{35} \dot{r}^3 \nu
+\frac{324}{7} \dot{r}^3 \nu^2
+\vek{v}^2
\left(
-\frac{18656}{105} \dot{r} \nu
-\frac{1116}{35} \dot{r} \nu^2
\right)
\right]\vek{n}
+\left[
-\frac{17308}{105} \dot{r}^2 \nu
-\frac{244}{21} \dot{r}^2 \nu^2
\right.
\nonumber\\ &
\left.
+\vek{v}^2
\left(
\frac{4394}{105} \nu
-\frac{16}{35} \nu^2
\right)
\right]\vek{v}
\bigg\}
+\frac{G^4 M^4}{r^5}
\left[
\left(
-\frac{152}{15} \dot{r} \nu
-\frac{632}{105} \dot{r} \nu^2
\right)\vek{n}
-\left(
\frac{386}{105} \nu
+\frac{16}{15} \nu^2
\right)\vek{v}
\right]
\Bigg\}
\label{eq:a12relativ}
\end{align}
with the notation
\begin{align}
\dot{r}=\left(\vek{n}\cdot\vek{v}\right)=\left(\vek{n}_{12}
\cdot\vek{v}_{12}\right)
\,.
\end{align}
This result corresponds to the generic equation of motion~(1.4) and
Eqs.~(2) in Ref.~\cite{IW95} by Iyer and Will (in the following
equations we take $G=1=c$):
\begin{align}
\frac{d^2 \vek{r}}{d t^2}=\,&
-\frac{M\,}{r^2}\vek{n}
+\frac{M}{r^2}
\bigg[
\vek{n}\left(A_{\rm 1PN}+A_{\rm 2PN}+A_{\rm 3PN}\right)
+\dot{r}\vek{v}\left(B_{\rm 1PN}+B_{\rm 2PN}+B_{\rm 3PN}\right)
\bigg]
\nonumber\\[1mm]
&+\frac{8}{5}\,\nu\,\frac{M}{r^2}\frac{M}{r}
\bigg[
\dot{r}\vek{n}\left(A_{\rm 2.5PN}+A_{\rm 3.5PN}\right)
-\vek{v}\left(B_{\rm 2.5PN}+B_{\rm 3.5PN}\right)
\bigg]
\end{align}
\end{widetext}
with [see Eqs.~(2.2a), (2.2b), (2.5) and (2.12) in Ref.~\cite{IW95}]
\begin{subequations}
\begin{align}
A_{\rm 1PN}=\,&-(1+3\,\nu)\vek{v}^2+\frac{3}{2}\nu\,\dot{r}^2+2(2+\nu)
\frac{M}{r}
\,,\\
B_{\rm 1PN}=\,&2 ( 2-\nu)
\,,\\
\label{eq:A25PNAllgemein}
A_{\rm 2.5PN}=\,&a_1 \vek{v}^2+a_2\frac{M}{r}+a_3\,\dot{r}^2
\,,\\
B_{\rm 2.5PN}=\,&b_1 \vek{v}^2+b_2\frac{M}{r}+b_3\,\dot{r}^2
\,,\\
A_{\rm 3.5PN}=\,&c_1 \vek{v}^4+c_2\vek{v}^2\frac{M}{r}+c_3\vek{v}^2\,\dot{r}^2
+c_4\,\dot{r}^2\frac{M}{r}
\nonumber\\
&+c_5\,\dot{r}^4+c_6\frac{M^2}{r^2}
\,,\\
\label{eq:B35PNAllgemein}
B_{\rm 3.5PN}=\,&d_1 \vek{v}^4+d_2\vek{v}^2\frac{M}{r}+d_3\vek{v}^2\,\dot{r}^2
+d_4\,\dot{r}^2\frac{M}{r}
\nonumber\\
&+d_5\,\dot{r}^4+d_6\frac{M^2}{r^2}
\,.
\end{align}
\end{subequations}
Note that we are not interested in the $2$PN and $3$PN terms in the present
work.

Iyer and Will establish in Refs.~\cite{IW93,IW95} that the energy
and angular momentum balances hold if and only if the coefficients 
$a_i, b_i, c_i$ and $d_i$ satisfies the following relations:
\begin{subequations}
\begin{align}
a_1&=3+3\,\beta
\,,&
a_2&=\frac{23}{3}+2\,\alpha-3\,\beta
\,,&
a_3&=-5\,\beta
\,, \\
b_1&=2+\alpha
\,,&
b_2&=2-\alpha
\,,&
b_3&=-3-3\,\alpha
\end{align}
\end{subequations}
for the $2.5$PN coefficients [see Eqs.~(2.11) of Ref.~\cite{IW95}], and
\begin{subequations}
\begin{align}
c_1=&\frac{1}{28}(117+132\,\nu)-\frac{3}{2}\beta(1-3\,\nu)+3\,\delta_2-3\,
\delta_6 
\,,\\
c_2=&-\frac{1}{42}(297-310\,\nu)-3\,\alpha(1-4\nu)-\frac{3}{2}\beta(7+13\,\nu)
\nonumber\\
&-2\,\delta_1-3\,\delta_2+3\,\delta_5+3\,\delta_6
\,,\\
c_3=&\frac{5}{28}(19-72\,\nu)+\frac{5}{2}\beta(1-3\,\nu)-5\,\delta_2
+5\,\delta_4+5\,\delta_6
\,,\\
c_4=&-\frac{1}{28}(687-368\,\nu)-6\,\alpha\,\nu+\frac{1}{2}\beta(54+17\,\nu)
\nonumber\\
&-2\,\delta_2-5\,\delta_4-6\,\delta_5
\,,\\
c_5=&-7\,\delta_4
\,,\\
c_6=&-\frac{1}{21}(1533+498\,\nu)-\alpha(14+9\,\nu)+3\,\beta(7+4\,\nu)
\nonumber\\
&-2\,\delta_3-3\,\delta_5
\,,\\
d_1=&-3(1-3\,\nu)-\frac{3}{2}\alpha(1-3\,\nu)-\delta_1
\,,\\
d_2=&-\frac{1}{84}(139+768\,\nu)-\frac{1}{2}\alpha(5+17\,\nu)+\delta_1-
\delta_3
\,,\\
d_3=&\frac{1}{28}(369-624\,\nu)+\frac{3}{2}(3\,\alpha+2\,\beta)(1-3\,\nu)
\nonumber\\
&+3\,\delta_1-3\,\delta_6
\,,\\
d_4=&\frac{1}{42}(295-335\,\nu)+\frac{1}{2}\alpha(38-11\,\nu)-3\,\beta(1
-3\,\nu)
\nonumber\\
&+2\,\delta_1+4\,\delta_3+3\,\delta_6
\,,\\
d_5=&\frac{5}{28}(19-72\,\nu)-5\,\beta(1-3\,\nu)+5\,\delta_6
\,,\\
d_6=&-\frac{1}{21}(634-66\,\nu)+\alpha(7+3\,\nu)+\delta_3
\end{align}
\end{subequations}
for the $3.5$PN coefficients [see Eqs.~(2.18) of
Ref.~\cite{IW95}].\footnote{In Ref.~\cite{IW95} the notation
$\varepsilon_5$ was used in place of $\delta_6$.}

The two parameters $\alpha$, $\beta$ at the $2.5$PN order and the six
coefficients $\delta_i$ at the $3.5$PN order correspond to the gauge
freedom and have no physical meaning. As mentioned in
Ref.~\cite{PatiWill2002}, at the $2.5$PN order, the values
\mbox{$\alpha=-1$}, \mbox{$\beta=0$} characterize the gauge of Damour and Deruelle
\cite{DamourDeruelle87A}, while \mbox{$\alpha=4$}, \mbox{$\beta=5$} correspond to the
so-called Burke-Thorne gauge [see for example \S36.11 of
\cite{GravitationMTW}] also used by Blanchet \cite{B93}.

It is then a non-trivial check to verify that the 18 coefficients $a_i$, $b_i$,
$c_i$ and $d_i$ parameterizing our $2.5$PN and $3.5$PN terms
[Eqs.~(\ref{eq:A25PNAllgemein})--(\ref{eq:B35PNAllgemein}) specialized
for Eq.~\eqref{eq:a12relativ}] yield a unique, self-consistent
solution for the 8 gauge coefficients. They are \cite{CK2002}:

for the $2.5$PN order:
\begin{subequations}
\begin{align}
\alpha=\frac{5}{3} \,, \quad\quad\quad \beta=3 \,;
\end{align}

for the $3.5$PN order:
\begin{align}
\delta_1=&\frac{41}{84}+\frac{677}{42}\nu\,,&
\delta_2=&-\frac{61}{14}-\frac{151}{14}\nu\,,
\\[3mm]
\delta_3=&\frac{583}{28}-\frac{157}{21}\nu\,,&
\delta_4=&\frac{115}{28}-\frac{15}{7}\nu\,,
\\[3mm]
\delta_5=&-\frac{961}{42}+\frac{16}{3}\nu\,,&
\delta_6=&-\frac{51}{14}-\frac{23}{14}\nu
\,.
\end{align}
\end{subequations}
\section{Discussion}

In this paper, we have computed PN reactive contributions to the ADM 
Hamiltonian for a pair of compact objects moving in general orbits,
using existing results available for the post-Newtonian $N$-body
Hamiltonian, computed by Jaranowski and
Sch\"afer~\cite{1997Jaranowski}. The two-body Hamiltonian is employed to
compute the first PN corrections to the instantaneous gravitational
energy loss associated with inspiraling compact binaries. We also derive
PN corrections to reactive equations of motion via Hamiltonian and
Euler-Lagrangian methods. The expressions for PN reactive
accelerations, computed in the ADM gauge, are consistent with those
of Iyer and Will. Employing the various reactive equations of motion
we also computed instantaneous gravitational wave luminosity for the
compact binary in general orbits. These luminosities are
in total agreement with their orbital averaged counterparts in the
case of quasi-elliptical orbital motion.
\begin{acknowledgments}
We thank P.~Jaranowski for discussions and useful checks.
We are grateful to A.~Gopakumar for a critical reading of
the manuscript. This work was supported by the EU Programme
``Improving the Human Research Potential and the Socio-Economic
Knowledge Base'' (Research Training Network Contract HPRN-CT-2000-00137)
and by the Deutsche Forschungsgemeinschaft (DFG) through SFB/TR7
Gravitationswellenastronomie.
\end{acknowledgments}
\appendix

\section{Hadamard Regularization}

Let $f$ be a smooth real-valued function defined on $\mathbb{R}^3$
deprived of a point $\vek{x}_0 \in \mathbb{R}^3$,
where it may be singular. We consider the family of auxiliary
functions \mbox{$f_\vek{n}(\varepsilon):=f\left(\vek{x}_0 +\varepsilon
\vek{n}\right)$}, labeled by unit vectors $\vek{n}$.
We expand $f_\vek{n}$ into Laurent series around $\varepsilon = 0$:
\be
\label{eq:laurent} f_\vek{n}(\varepsilon) = \sum\limits_{m =
-N}^{\infty} a_m(\vek{n})\,\varepsilon^m\,.
\ee
The coefficients $a_m$ of this expansion depend on the unit vector
$\vek{n}$. We define the regularized value of the function $f$ at
$\vek{x}_0$ as the coefficient of $\varepsilon^0$ in the expansion
\eqref{eq:laurent} averaged over all directions:
\be
\label{eq:HPFdef} f_{\rm reg}\left(\vek{x}_0\right)
:=\frac{1}{4\pi}\oint \! a_0(\vek{n}) \,d\Omega
\,,
\ee
where $d\Omega$ is the elementary solid angle viewed from the point
$\vek{x}_0$.
This procedure is called Hadamard regularization. To calculate the
integrals in Eq.~\eqref{eq:HPFdef} we used the following equations
\begin{subequations}
\begin{align}
\frac{1}{4\pi}\oint \! n^{i_1} n^{i_2} \ldots n^{i_l} \,d\Omega
&=0 \,,\quad \quad\quad \text{for odd $l$}\,,
\\[1mm]
\frac{1}{4\pi}\oint \!n^{i_1} n^{i_2} \ldots n^{i_{2p}} \,d\Omega&=
\frac{1}{(2p+1)!!}
\underbrace{\delta^{\{i_1 i_2} \ldots \,\delta^{i_{2p-1}
i_{2p}\}}}_{(2p-1)!! \quad {\rm terms}}
\,,
\label{eq:RingintegralGerade}
\end{align}
\end{subequations}
with $(2p+1)!!=1\cdot3\cdot5\cdot\ldots\cdot(2p+1)$. The
curly braces in Eq.~(\ref{eq:RingintegralGerade}) mean the symmetric
combination of the indices:
\be
\delta^{\{ij}\,\delta^{kl\}}=\underbrace{\delta^{ij}\delta^{kl}+\delta^{ik}
\delta^{jl}+\delta^{il}\delta^{jk}}_{3 !!\quad {\rm terms}}\,.
\ee

During the identification process of the primed and unprimed variables,
$\vek{x}_{1'} \rightarrow \vek{x}_{1}$ and
$\vek{x}_{2'} \rightarrow \vek{x}_{2}$, in the
calculation of the derivation of the Hamiltonians given by the
Eqs.~\eqref{eq:25res} and \eqref{eq:35res}, some singularities appear.
The purpose of the regularization is the elimination of
these singularities. It is a matter of choice whether we apply first the
regularization for index $1'$ and then for index $2'$, or the other way
around. 

As an example, we shall apply Hadamard regularization procedure on the term
\begin{multline}
\frac{\left( \vek{n}_{11'} \cdot \vek{n}_{22'} \right)
\left( \vek{n}_{11'} \cdot \vek{p}_{1} \right)
\left( \vek{n}_{22'} \cdot \vek{p}_{2} \right)}{r_{11'}^2 r_{1'2'}^3}
\\ =n_{11'}^i n_{22'}^i n_{11'}^j p_{1}^j n_{22'}^k p_{2}^k
r_{11'}^{-2} r_{1'2'}^{-3}
\label{eq:ExampleTerm}
\,.
\end{multline}
We intend to identify index $1$ with $1'$ as a first step.
We start by calculating the limit for
\mbox{$\varrho_1^i = - r_{11'}^i \rightarrow 0$} after the replacement:
\begin{align}
r_{1'2'}^i
&= r_{12'}^i - r_{11'}^i = r_{12'}^i + \varrho_1^i
\,.
\end{align}
in Eq.~\eqref{eq:ExampleTerm}. To this purpose, we expand
\begin{align}
r_{1'2'}^m &= f\left(r_{12'}^i + \varrho_1^i\right)
\nonumber \\
&=\sum\limits_{l=0}^{+\infty} \frac{\varrho_1^L}{l !} \pa_L r_{12'}^m
=\sum\limits_{l=0}^{+\infty} \frac{(-1)^l}{l !} r_{11'}^l
n_{11'}^L \pa_L r_{12'}^m
\,.
\end{align}
We now perform the first angular integration over the terms of
Eq.~\eqref{eq:ExampleTerm} depending on the variable labeled by $1'$.
Here is $m=-3$ and $l=2$.
\begin{align}
&\frac{1}{4\pi}\oint \! n_{11'}^i n_{11'}^j r_{11'}^{-2} r_{1'2'}^{-3}
\,d\Omega_{11'} 
\nonumber \\
&\quad =\sum\limits_{l=0}^{+\infty} \frac{(-1)^l}{l !} r_{11'}^{l-2}
\pa_{(12')L} \,r_{12'}^{-3}
\frac{1}{4\pi}\oint \! n_{11'}^i n_{11'}^j n_{11'}^L \,d\Omega_{11'}
\nonumber \\
&\quad =\frac{1}{2} \pa_{(12')ab} \,r_{12'}^{-3}
\frac{1}{4\pi} \oint \! n_{11'}^i n_{11'}^j n_{11'}^a n_{11'}^b \,d\Omega_{11'}
\nonumber \\
&\quad =\frac{1}{2} \pa_{(12')ab} \,r_{12'}^{-3} \frac{1}{15}
\left(
\delta^{ij} \delta^{ab}
+\delta^{ia} \delta^{jb}
+\delta^{ib} \delta^{ja}
\right)
\nonumber \\
&\quad =\frac{1}{30} \left( 2 \pa_{(12')ij} \,r_{12'}^{-3}
+\delta^{ij} \Delta_{(12')} \,r_{12'}^{-3} \right)
\,.
\end{align}
By using
\begin{subequations}
\begin{align}
\pa_{ij} r^m &= \left[ m(m-2) n^i n^j + m \delta^{ij} \right] r^{m-2}
\,, \\
\Delta r^m &= m(m+1) r^{m-2}
\,,
\end{align}
\end{subequations}
we get
\be
\label{eq:FirstResultofAngularIntg}
\frac{1}{4\pi}\oint \! n_{11'}^i n_{11'}^j r_{11'}^{-2} r_{1'2'}^{-3}
\,d\Omega_{11'}=\frac{n_{12'}^i n_{12'}^j}{r_{12'}^5}
\,.
\ee
In further computations we replace the terms on the right-hand side of
Eq.~\eqref{eq:FirstResultofAngularIntg} with 
\begin{subequations}
\label{eq:r12'UNDn12'}
\begin{align}
r_{12'}^i &= r_{12}^i + r_{22'}^i =r_{12}^i + \varrho_2^i \,,
\label{eq:r12'}
\\
n_{12'}^i &= \frac{r_{12}}{r_{12'}} n_{12}^i
+\frac{r_{22'}}{r_{12'}}n_{22'}^i
=\frac{r_{12}}{r_{12'}} n_{12}^i
+\frac{\varrho_2}{r_{12'}}n_{22'}^i
\label{eq:n12'}
\,.
\end{align}
\end{subequations}
The second step will consist in identifying index $2$ with $2'$.
We calculate the limit for $\varrho_2^i =r_{22'}^i \rightarrow 0$.
The second angular integration over the finite part terms (\emph{i.e.}
the terms that are of order zero in power of $\varrho_2^i$) depending on the
variable labeled by $2'$ can be performed immediately because there are
no singularities in Eqs.~\eqref{eq:r12'UNDn12'} for
\mbox{$\varrho_2^i \rightarrow 0$}:
\begin{multline}
\frac{n_{12}^i n_{12}^j p_1^j p_2^k }{r_{12}^5}
\left( \frac{1}{4 \pi} \oint \! n_{22'}^i n_{22'}^k
\,d\Omega_{22'} \right)
\\
= \frac{n_{12}^i n_{12}^j p_1^j p_2^k }{r_{12}^5}
\left( \frac{1}{3} \delta^{ik} \right)
\,.
\end{multline}
The result is
\begin{multline}
\left[
\frac{\left( \vek{n}_{11'} \cdot \vek{n}_{22'} \right)
\left( \vek{n}_{11'} \cdot \vek{p}_{1} \right)
\left( \vek{n}_{22'} \cdot \vek{p}_{2} \right)}{r_{11'}^2 r_{1'2'}^3}
\right]_{\left[\onunder{\vek{x}_{1'} \rightarrow \vek{x}_{1}}{
\vek{x}_{2'} \rightarrow \vek{x}_{2}}\right]}
\\
=\frac{\left( \vek{n}_{12} \cdot \vek{p}_1 \right)
\left(\vek{n}_{12} \cdot \vek{p}_2 \right)}{3 r_{12}^5}
\,.
\end{multline}

\bibliographystyle{apsrev}

\end{document}